\documentclass[10pt,journal,compsoc]{IEEEtran}
\ifCLASSOPTIONcompsoc
  \usepackage[nocompress]{cite}
\else
  \usepackage{cite}
\fi
\ifCLASSINFOpdf
\else
\fi
\usepackage{booktabs}
\usepackage{threeparttable}
\usepackage{makecell}
\usepackage{enumitem}
\usepackage{xspace}
\usepackage{verbatim}
\usepackage{tikz}
\usepackage{colortbl}
\usepackage{hyperref}
\usepackage{collectbox}
\usepackage{multirow}
\usepackage{amssymb}
\usepackage{amsthm}
\usepackage{algpseudocode}
\usepackage{algorithm}

\theoremstyle{definition}

\usepackage[caption=false]{subfig}

\usepackage{amsmath}

\usepackage{pifont}

\newcommand{\header}[1]{\vspace{-1.5mm}\par\medskip\noindent\textbf{#1:}}
\usepackage{circledsteps}
\definecolor{foo}{HTML}{4371C2}

\usepackage[nomessages]{fp}
\newcommand{\FPuse}[1]{\FPeval{\result}{#1}{\result}}
\newcommand{\tableRow}[3]{
	#1 & #2 & #3 &
	\ifthenelse{\equal{#2}{0}}{$100$}{$\FPuse{round(#1 / (#1 + #2) * 100, 2)}$}
	&
	\ifthenelse{\equal{#3}{0}}{$100$}{$\FPuse{round(#1 / (#1 + #3) * 100, 2)}$}
	&
	$\FPuse{round(2 * #1 / (2* #1 + #2 + #3) * 100, 2)}$
}

\newcommand{\tableRowBold}[3]{
	#1 & #2 & #3 &
	\ifthenelse{\equal{#2}{0}}{\boldmath$100$}{\boldmath$\FPuse{round(#1 / (#1 + #2) * 100, 2)}$}
	&
	\ifthenelse{\equal{#3}{0}}{\boldmath$100$}{\boldmath$\FPuse{round(#1 / (#1 + #3) * 100, 2)}$}
	&
	\boldmath$\FPuse{round(2 * #1 / (2* #1 + #2 + #3) * 100, 2)}$
}

\newcommand\sbar[3]{
	#1 & #2 & #3 &
	\ifthenelse{\equal{#2}{0}}{$100$}{$\FPuse{round(#1 / (#1 + #2) * 100, 2)}$} &
	\ifthenelse{\equal{#3}{0}}{$100$}{$\FPuse{round(#1 / (#1 + #3) * 100, 2)}$} 
}

\usepackage{tcolorbox}
\definecolor{findingsbox-bg-color}{gray}{0.90}
\newtcbox{\findingsbox}{colback=findingsbox-bg-color, boxrule=0.2pt, arc=2pt, boxsep=0pt, left=5pt, right=5pt, top=5pt, bottom=3pt}
\newcommand{\findings}[2] {
	\noindent
	\findingsbox{
		\begin{minipage}{.95\linewidth}
			\textbf{#1}: #2
		\end{minipage}
	}
}

\begin{document}
\title{Refactoring-aware Block Tracking\\in Commit History}

\author{Mohammed Tayeeb Hasan, Nikolaos~Tsantalis,~\IEEEmembership{Senior~Member,~IEEE} and Pouria Alikhanifard
}

\IEEEtitleabstractindextext{%
	\begin{abstract} 
Tracking statements in the commit history of a project is in many cases useful for supporting various software maintenance, comprehension, and evolution tasks. 
A high level of accuracy can facilitate the adoption of code tracking tools by developers and researchers. 
To this end, we propose CodeTracker, a refactoring-aware tool that can generate the commit change history for code blocks.
To evaluate its accuracy, we created an oracle with the change history of 1,280 code blocks found within 200 methods from 20 popular open-source project repositories.
Moreover, we created a baseline based on the current state-of-the-art Abstract Syntax Tree diff tool, namely GumTree 3.0, in order to compare the accuracy and execution time. Our experiments have shown that CodeTracker has a considerably higher precision/recall and faster execution time than the GumTree-based baseline, and can extract the complete change history of a code block with a precision and recall of 99.5\% within 3.6 seconds on average.
	\end{abstract}
	\begin{IEEEkeywords}
		Commit change history, Refactoring-aware source code tracking, Change oracle
	\end{IEEEkeywords}
}

\maketitle

\IEEEdisplaynontitleabstractindextext

\section{Introduction}
Developers routinely track code snippets in the commit history to facilitate various software engineering tasks.
Codoban et al.~\cite{Codoban:ICSME:2015} surveyed 217 developers to find the motivations behind examining software history.
The most common reasons are to
a) recover the rationale behind a snippet of code, 
b) find the commits that introduced a bug,
c) find who are the knowledgeable peers on certain modules and patterns,
d) reverse engineer requirements from code,
e) keep up with how the code state evolves,
f) apply changes from other branches into the main branch.

\subsection{Motivation}
\header{Developers} Grund et al.~\cite{Grund:2021} conducted a survey with 42 professional software developers
and found that they prefer source code history information at the method/function and class level rather than the file level.
Moreover, the tools used by the developers to inspect code history, such as \texttt{git log} and IntelliJ's history feature,
are unable to find the commit that introduced a method or deal with complex structural changes (e.g., method moves).
Codoban et al. \cite{Codoban:ICSME:2015} surveyed 217 developers and
found that 85\% of them consider software history important
to their development activities and 61\% need to refer to
history at least several times a day.
The surveyed developers expressed some challenges regarding
the usability of existing tools, such as their inability to detect file moves and renames,
and their difficult configuration (e.g., setting up \texttt{git bisect} to find the commit that introduced a bug).
LaToza and Myers~\cite{LaToza:Hard-to-Answer-Questions} surveyed 179 professional software developers at Microsoft and asked them to list hard-to-answer questions that they had recently asked about code.
Among the collected responses, developers asked about ``\textit{Where was this variable last changed?}'' when debugging,
``\textit{When, how, by whom, and why was this code changed or inserted?}'' when they want to find the code’s creation in history to understand its context and motivation,
and finally ``\textit{How has it changed over time?}'' when they want to know the entire history of a block of code, rather than its most recent change.
Fritz and Murphy~\cite{Fritz:2010} surveyed eleven professional software developers to find questions that developers ask, but have no resources that can help to answer them. Out of the 78 recorded questions, 20 code-specific questions were highlighted in their paper. Among them, developers were interested in knowing who originally wrote a piece of code and who modified it last.
Ko et al.~\cite{Ko:2007} surveyed seventeen software developers and logged their activities minute by minute. They found that developers wanted to know more about \textit{“Why was this code implemented this way?”}, so as to derive historical reasoning for its current implementation. Another interesting point made in this paper is that during bug fixes, developers need specific code change history to analyze whether the error was anticipated by the designer and explicitly ignored or whether it was overlooked. Thus, having block-level source code history can speed up bug-fixing efforts in cases where bugs are known to be present in a specific block rather than the entire method.

These findings motivate the need for developing tools that can track change history at a more \textit{fined-grained level}, focusing on specific program elements, such as methods/functions, variables, and code blocks. Moreover, such tools should be \textit{refactoring-aware} and support complex structural changes that move the tracked program element to a distant location within the same or a different file.

\header{Researchers} Accurate code snippet tracking is also essential in many areas of software engineering research.
Alencar da Costa et al.~\cite{AlencarDaCosta:2016} pointed out that
bug-inducing analysis algorithms (e.g., SZZ~\cite{Sliwerski:2005,Kim:2006,Williams:2008}) suffer from broken historical links due to file moves and renames.
This further affects the results of defect prediction techniques and empirical studies investigating the characteristics of bug-introducing changes, which rely on the original SZZ algorithm or its variants~\cite{Rosa:ICSE:2021}.
Jian et al.~\cite{BugBuilder} claimed that
fully automated construction of bug repositories by mining bug-fixing commits from version control systems often results in inaccurate patches that contain many bug-irrelevant changes, such as overlapping refactorings and non-essential changes.
Inaccurate big-fixing patches negatively affect several research areas related to bugs, such as fault localization, program repair, and software testing~\cite{BugBuilder}.
Shen et al.~\cite{Shen:OOPSLA:2019} showed that automatic source code merging tools
often fail to track the changed program elements correctly due to overlapping refactoring operations, and thus are unable to perform the auto-merging.
The automatic migration of client software to newer library and framework versions requires tracking the updated API program elements (i.e., methods and fields) from the source to the target version, extracting changes in the API signatures, and adapting the API references in the client's code~\cite{Dig:2006,Kapur:OOPSLA:2010,Cossette:FSE:2012}.
API program element tracking has been performed both at commit level~\cite{Brito:SANER:2018,Brito:APIDiff:2018} and release level~\cite{McDonnell:ICSM:2013,Mahmoudi:MSR:2018}. 
However, fine-grained program element tracking at the commit level may be more accurate than release level~\cite{Brito:EMSE:2020},
as comparing two releases involves significantly more noise from overlapping changes performed in all commits between the two releases.

The inherent limitations of the line-based text diff and blame tools, which are predominantly used in the aforementioned software engineering tasks, motivated researchers to develop techniques for tracking more accurately statements/lines \cite{10.1109/MSR.2007.14, 10.1109/MS.2009.26, 10.1145/1368088.1368091, 10.1007/s10515-006-0002-0, 10.1109/SCAM.2009.19, 10.1145/2393596.2393646}, as well as program elements, such as methods/functions, attributes and classes~\cite{Grund:2021,Historage,FinerGit,Kim:WCRE:2005,Godfrey:TSE:2005,Steidl:MSR:2014,Tempura:ICSE:2015,Hora:ICSE:2018}, in the commit history of software repositories. These techniques deal with changes that modify the name/signature or location of a program element and can cause a split in its history. Hora et al.~\cite{Hora:ICSE:2018} found that 25\% of classes and methods have at least one \textit{untracked} change (i.e., move, rename, extract, inline refactoring) in their histories.
Despite the significant accuracy improvements brought by the aforementioned tools, they still have some major limitations, which are discussed in Section~\ref{sec:related-work}.

\header{Why is tracking code blocks more challenging than other program elements?}
Methods, fields, and types have a unique signature, as the compiler does not allow to have multiple program elements with the same signature within the same scope.
In contrast, code blocks are structures without a unique signature.
Moreover, it is possible to have multiple textually identical blocks within the body of the same method. Therefore, using simple textual similarity metrics to match code blocks would potentially result in multiple matching candidates.
Theoretically, the nesting depth of a code block along with its index in the parent list of statements can be used to uniquely identify the location of this code block within its container method. However, such a location signature is sensitive to \textit{control re-structuring}, \textit{statement re-ordering}, \textit{refactorings}, and \textit{addition/deletion of code blocks} within a method. Therefore, in order to reliably track a code block between two versions, we need a statement mapping algorithm that can robustly handle the aforementioned edit operations, which change the structure of a method.

For this reason, we decided to depend on RefactoringMiner 3.0~\cite{Tsantalis:TSE:2020:RefactoringMiner2.0, Alikhanifard:TOSEM:2024:RefactoringMiner3.0} to build our code block tracking solution.
RefactoringMiner is a mature tool
maintained over a period of 8 years by members of our Refactoring Research Group at Concordia University. It currently
supports the detection of 100 different refactoring types and API changes and has been established as the tool
with the highest accuracy and fastest execution time among competitive tools.
Under the hood, RefactoringMiner applies a statement mapping process. The mapped statements and the AST node replacements found within the mapped statements are then used to infer refactoring operations.

Our solution has been designed for imperative object-oriented code and its current implementation supports Java code.
We consider as blocks all AST statements that can contain nested statements within their body.
The complete list of such statements is shown in Table~\ref{table:block-types}.
Our solution does not support the tracking of a sequence of statements that are not nested under a control structure, as each statement could have its own individual change history in this case.
However, if this sequence of statements is placed within an AST block (i.e., opening and closing curly brackets \texttt{\{...\}} without a control statement)
our solution can track the parent block statement, as RefactoringMiner is capable of matching block statements.

\subsection{Contributions}
This work has the following novel contributions:
\begin{enumerate}[leftmargin=*]
	\item We create a new oracle with the change history
	of 1,280 code blocks declared within 200 methods from 20 popular open-source project repositories (10 methods from each repository).
	To the best of our knowledge, this is the first oracle in the literature including change history for code blocks. Moreover, it complements existing oracles that include the change history of the same 200 methods~\cite{Grund:2021, Jodavi:2022} and the variables declared within the body of these 200 methods~\cite{Jodavi:2022}.
	\item We implement CodeTracker 2.0 to support the tracking of code blocks. 
	To the best of our knowledge, CodeTracker is the only tool that can construct the change history of code blocks in a fully refactoring-aware fashion.
	Our tool can track code blocks transformed to a different AST type (e.g., \texttt{for} changed to \texttt{while} loop),
	and supports forks in the evolution history of a block occurring when two or more different blocks are merged into one.
	CodeTracker is publicly available on GitHub~\cite{CodeTrackerOracle} and Maven central repository~\cite{codetracker-maven}.
	\item We create a Chrome browser extension~\cite{codetracker-visualizer} that can be used to navigate and inspect the change history of methods, variables, and code blocks directly on GitHub. The same extension was used to create and validate our oracle of block changes.
	\item We develop a baseline based on the GumTree AST diff tool~\cite{gumtree, gumtree-hyperparameter-optimization} to evaluate and compare the accuracy of our tool.
	Moreover, we conduct experiments showing that CodeTracker has considerably higher precision/recall and faster execution time than the baseline tool based on GumTree.
	\item We further improve the performance of CodeTracker in Step 5 (i.e., the most time-consuming step of the approach) by removing from the partial source code models for the parent and child commits all pairs of method declarations that are identical within files having the same file path.
\end{enumerate}
\section{Approach}
This section presents our approach for modeling and reconstructing the changes applied on a code block in the commit history of a project.

\subsection{Code Block Identifier}
\label{sec:code-block-identifier}

Jodavi and Tsantalis \cite{Jodavi:2022} defined each code element $e$ to be uniquely identified in the commit history of a software repository with the following tuple:

\begin{equation}
	\label{equ:1}
	I_e=(V_e, CON_e, SIG_e)
\end{equation}
where $V_e$ is the version of $e$ corresponding to the SHA-1 git commit ID in which a change took place on code element $e$, $CON_e$ is the signature of the container to which $e$ belongs, and $SIG_e$ is the signature of $e$.

Building upon this design, the identifier for a code block $b$ is defined as follows:

\begin{equation}
	\label{equ:2}
	I_b=(V_b, CON_b, SIG_b)
\end{equation}
where $V_b$ is the version of $b$ corresponding to the commit ID in which $b$ changed. 
The container of $b$ is the tuple: 
\begin{equation}
	\label{equ:3}
	CON_b = (CON_{M_b}, SIG_{M_b})
\end{equation}
where $M_b$ is the method declaration in which $b$ is declared, and $CON_{M_b}$ and $SIG_{M_b}$ are the container and signature of $M_b$, respectively.

The container of a method declaration $M$ is the tuple:
\begin{equation}
	\label{equ:4}
	CON_M = (CON_{C_M}, SIG_{C_M})
\end{equation}
where $C_M$ is the type declaration to which $M$ belongs, and $CON_{C_M}$ and $SIG_{C_M}$ are the container and signature of $C_M$, respectively, as defined by Jodavi and Tsantalis \cite{Jodavi:2022}.
The container of a type declaration $C$ is the tuple: 
\begin{equation}
	\label{equ:5}
	CON_C = (SRC_C, PKG_C)
\end{equation}
where $SRC_C$ is the source folder path and $PKG_C$ is the package name to which $C$ belongs.

Finally, the signature of code block $b$ is the tuple:
\begin{equation}
	\label{equ:7}
	SIG_b = (T_b, SIG_{p_b}, SIG_{body})
\end{equation}
where $T_b$, is the block type (e.g., \texttt{for}, \texttt{if}, \texttt{try}, \texttt{switch}), $SIG_{p_b}$ is the signature of $b$'s parent statement, and $SIG_{body}$ is the signature of $b$'s body, which is essentially the hash value of the code inside $b$'s body.

The signature of the parent statement $p_b$ has a recursive definition as shown in the tuple:
\begin{equation}
	\label{equ:6}
	SIG_{p_b} = (SIG_{{p\prime}_p}, T_p, I_{p_b}) 
\end{equation}
where $p\prime$ is the parent of $p$, $T_p$ is the statement type of $p$, and $I_{p_b}$ is the index of $b$ in $p$'s list of statements, respectively.
If $p$ corresponds to the body of $M_b$ (i.e., the method in which $b$ is declared), then $p\prime$ is \texttt{null} and $SIG_{{p\prime}_p}$ is an empty string.

This information is necessary to create a unique identifier for each code block, as there may exist multiple blocks within a method that are textually identical, but have a different location in the method's control and execution flow structure.

\subsection{Block Tracking Process}
\label{sec:block-tracking-process}
Our solution relies on the statement mappings generated by RefactoringMiner 3.0~\cite{Tsantalis:TSE:2020:RefactoringMiner2.0, Alikhanifard:TOSEM:2024:RefactoringMiner3.0} to track a code block in the commit history of a project, and report all changes performed on it, even if the code block itself or its parent container has been refactored.
Despite the fast execution time of RefactoringMiner (44 ms on median and 253 ms on average per commit), running it on the entire commit history of the project is computationally inefficient, as the tracked program element is changing in a relatively small subset of commits, and furthermore, it is not always necessary to analyze all modified files in a commit to track a single program element, especially in large commits involving thousands of modified files.
Therefore, we developed some heuristics and extended RefactoringMiner to perform \textit{partial} and \textit{incremental} commit analysis.

\header{Input} CodeTracker takes as input a Git repository URL, a starting commit SHA-1 ID (or HEAD by default), the file path containing the code block of interest, the type of the code block (e.g., \texttt{for}, \texttt{if}, \texttt{while}, \texttt{try}, \texttt{catch}), and the start line number of the code block in the file.

\header{Output} The output is a graph, where the nodes represent code elements with their unique identifiers, and the list of changes between two nodes is attached to the edge connecting them. The change history is returned in the form of a graph due to the possibility of forks. A fork occurs when two or more different blocks are merged into one.
For example, two or more \texttt{catch} blocks could be merged into a single \texttt{catch} block using the union type feature of Java for the handled exception types, e.g., \texttt{catch(ClassNotFoundException | IllegalAccessException ex)}.
Another example is the extraction of two or more duplicated code blocks from the same or different methods into a single commonly used method (i.e., \textsc{Extract Method} refactoring).
The detection of forks is possible because RefactoringMiner 3.0 supports a novel source code diff feature, namely \textit{multi-mappings} (i.e., the case where a statement from the child commit has more than one corresponding statement in the parent commit, and vice versa).
The change history of a code block starts from the commit provided as input and goes all the way back to its introduction commit. Therefore, by traversing the graph from the start node, we can visualize the changes that took place in each commit, and since the graph can contain forks, every block that was potentially merged to the tracked block can also be traced back to its introduction commit.

\begin{figure}[ht]
    \centering
    \vspace{-3mm}
    \includegraphics[width=\linewidth]{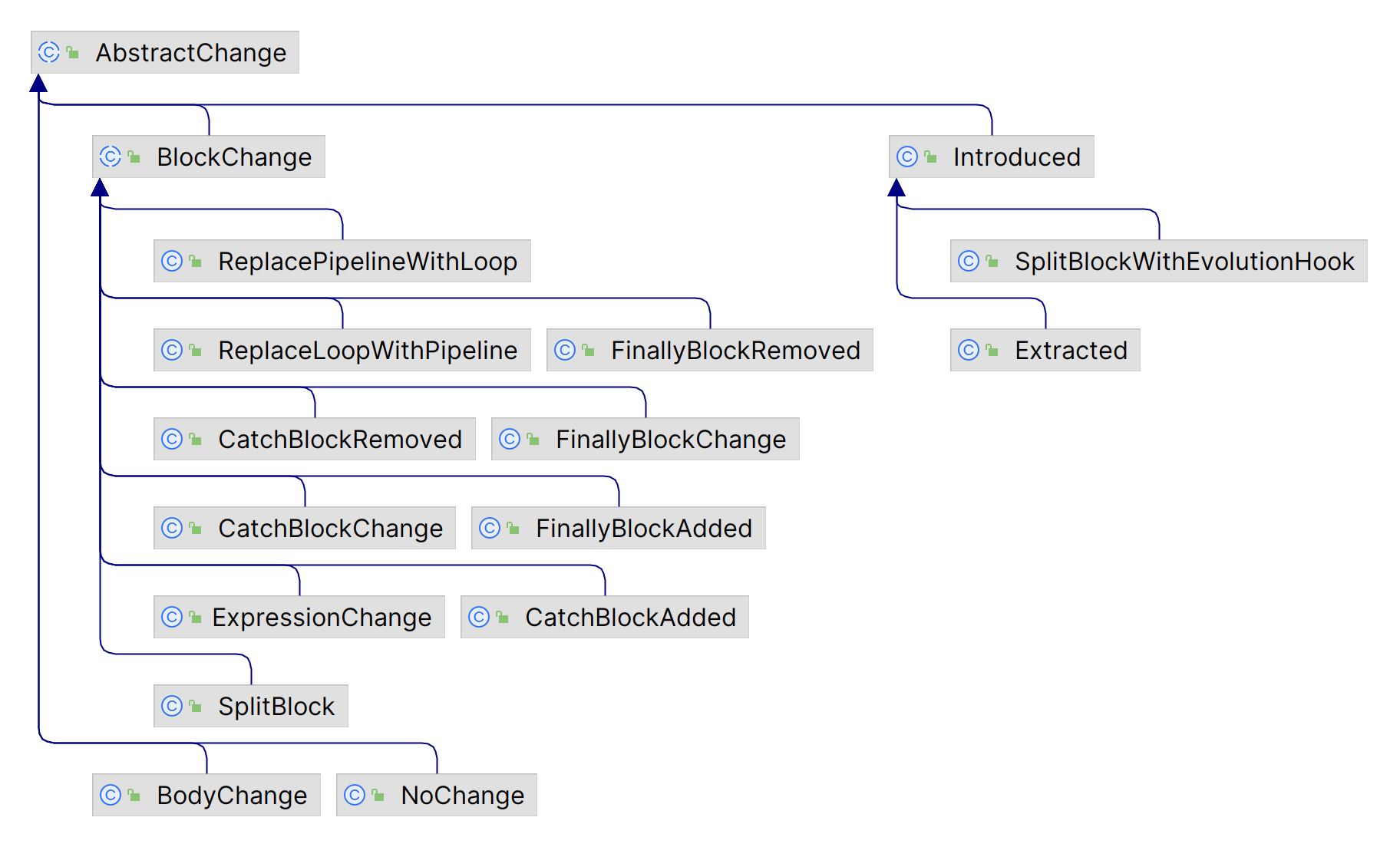}
    \vspace{-7mm}
    \caption{Hierarchy of supported change kinds for code blocks.}
    \vspace{-3mm}
    \label{fig:blockChangeTypes}
\end{figure}

Figure \ref{fig:blockChangeTypes} shows the change type hierarchy supported by CodeTracker for code blocks.
\textsc{Body} and \textsc{Expression} changes can be considered as the ``common denominator'' for all block types, as all of them have a body, and some of them have expressions within parentheses.
All other changes are specific to certain block types.
For example, all changes related to \texttt{catch} and \texttt{finally} blocks are applicable only for \texttt{try} blocks,
and we considered them, because such changes are not taking place within the \texttt{try} body or the \texttt{try} resource expressions.
The \textsc{Replace Pipeline with Loop} change \cite{Fowler:2018} is particularly interesting, because this is the only transformation that converts a block statement (i.e., a loop statement with a body containing nested statements) to a leaf statement (i.e., a statement without a body containing nested statements), as a pipeline is essentially a chain of Java Stream API calls.
The reverse change is also supported (i.e., \textsc{Replace Loop with Pipeline}).
These two changes are significant, because if not supported, then we may miss a large portion of the change history of a block, since many projects have migrated traditional \texttt{for} loops with nested conditional logic to the Stream API, and many IDEs offer refactoring support to automate such migrations \cite{Mazinanian:2017:UUL:3152284.3133909}.
Finally, \textsc{Split Block} typically occurs when the conditionals of an \texttt{if} statement are split into two or more separate \texttt{if} statements, which are nested within each other (if the original conditionals are combined with an \texttt{\&\&} operator), or are sequentially executed (if the original conditionals are combined with an \texttt{||} operator).
In such case, instead of considering the separated \texttt{if} statements as newly added blocks in the child commit, we resume the tracking process for the original \texttt{if} statement in the parent commit.
 

One interesting feature to point out is that the block tracking process also supports the transformation of blocks from one type to another. For example, in one case that we found, a \texttt{switch} statement was used to replace a rather cumbersome \texttt{if-else-if} ladder and then add a few extra cases \cite{switch-to-if}. We support continuous tracking in such instances, as the \texttt{switch} cases are mapped to the corresponding \texttt{if} conditionals, and the evolution chain continues. 
The complete list of such transformations includes (the inverse transformation of all these cases is also supported):
\begin{enumerate}
	\item \texttt{if-else-if} to \texttt{switch} cases
	\item \texttt{if} to \texttt{while} loop
	\item iterator-based \texttt{while} loop to \texttt{enhanced-for} loop
	\item \texttt{for} loop to \texttt{while} loop
	\item \texttt{for} loop to \texttt{forEach} pipeline
	\item \texttt{for} loop to \texttt{if}
	\item \texttt{try} block to \texttt{try-with-resources} block
	\item \texttt{try} block to \texttt{synchronized} block
	\item \texttt{catch} block to \texttt{finally} block
\end{enumerate}

It should be noted that the change types \textsc{Split Block with Evolution Hook} and \textsc{Extracted} shown in Figure \ref{fig:blockChangeTypes}, serve as \textit{evolution hooks}, a concept introduced in our previous work~\cite{Jodavi:2022}, allowing to pause the change history when a block is split or extracted from another method, respectively, and leaving the option to the user of our tool to attach on demand the remaining evolution sub-graph if needed.
This design choice allows us to avoid computing additional change history, which might not be needed by the user, but at the same time inform the user about the opportunity to further explore the change history, if and when desired.
However, in the current implementation, we automatically continue tracking the change history when a block is split or extracted from another method, and thus these two types are not instantiated and reported.
We consider code blocks as independent program elements, which can be moved between methods in the same file or even different files with refactorings, such \textsc{Extract}, \textsc{Inline}, \textsc{Split} and \textsc{Merge Method}, and thus their evolution history should not be interrupted when they are relocated to a new container.

An overview of the tracking process is shown in Figure \ref{fig:workflow} and
consists of the following steps:

\begin{figure}
    \centering
    \includegraphics[width=\linewidth]{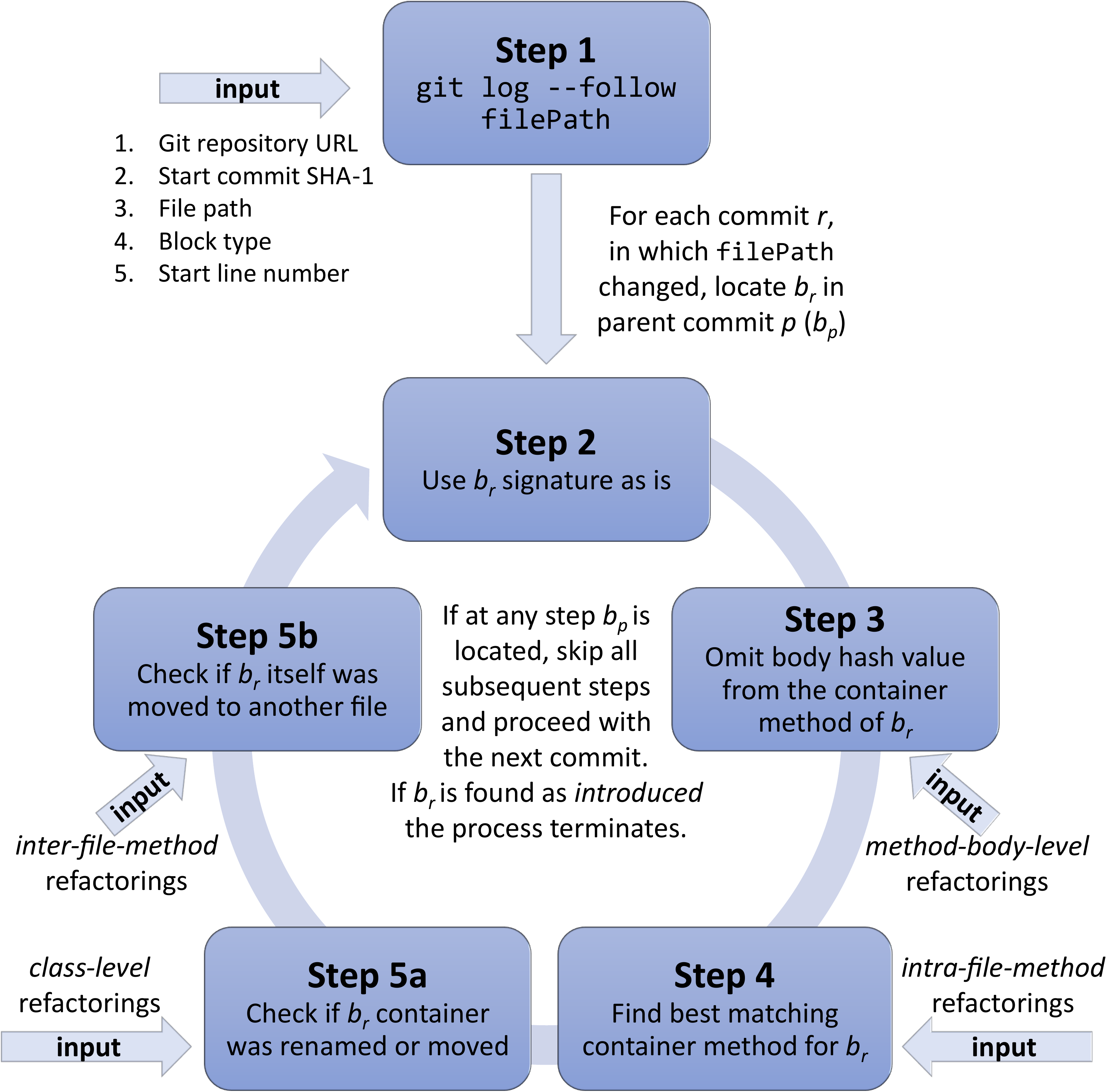}
    \vspace{-7mm}
    \caption{Overview of the block tracking process steps.}
    \vspace{-3mm}
    \label{fig:workflow}
\end{figure}

\header{\textsc{Step 1}. Retrieve Git history for the specified file path}

As a starting point, we locate the code block of interest (denoted by $b$) within the file and the start commit specified as input by the user.
If CodeTracker is not able to locate a valid code block with the specified input parameters, it throws a \texttt{CodeElementNotFoundException} and terminates the tracking process.

If a valid code block is found, we then retrieve the git history of the specified repository and obtain all the commits in which the file had undergone a change. We collate these commits and process them into the next step, ignoring the rest of the commits that do not include modifications for the specified file. 
This step avoids iteratively processing each commit in the repository. 
The command used for this process is \texttt{git log --follow filePath}, and by using the \texttt{follow} flag, we obtain the commits in which \texttt{filePath} is moved or renamed as well.

\header{\textsc{Step 2}. Check if the container method is unchanged} 

After obtaining the set of commits in which the file initially containing $b$ is modified, we iterate through each of these commits and construct a partial source code model for the file containing the block in the current commit $r$ and the same file in the parent commit $p$, respectively. 
Alongside the partial source code models, we also construct the signatures ($SIG_{M_r}$, $SIG_{b_r}$) and containers ($CON_{M_r}$, $CON_{b_r}$) of the method containing the block and the block itself in commit $r$, as illustrated in Section \ref{sec:code-block-identifier}. 
We then look into $p$’s model and search for a method with the same signature as $SIG_{M_r}$. 
If we do find a match, it means that the container method has remained unchanged, and thus so has the code block contained inside its body. 
As a result, it is possible to find a block $b_p$ within $M_p$ that has the same signature as $SIG_{b_r}$, excluding the version number.
We then link the type declaration containers ($CON_{M_p}$, $CON_{M_r}$), method declarations ($M_p$, $M_r$) and blocks ($b_p$, $b_r$) to each other and continue the tracking process. 
If a match is not found, we move on to \textbf{\textsc{Step 3}}.

\header{\textsc{Step 3}. Check if the container method body changed} 

Reaching this step would indicate that the container method $M_r$ has undergone some change in child commit $r$. Therefore, in this step, we check if the method remains in the same file in the parent commit $p$ and has only undergone a change in its body, that caused a change in its identifier. We do this by relaxing the method identifier when searching for a match. More specifically, by omitting the method body’s hash value, $SIG_{body}$, we can now compare to see if there is a method that matches this identifier in commit $p$. If we do find a match, this would indicate that the body of the method has changed, but its signature (i.e., method name, parameter type list, return type) remained the same. In turn, the block contained inside the body of this method could potentially have also changed. 

At this point, we execute RefactoringMiner on the partial source code models for commits $r$ and $p$ constructed in \textbf{\textsc{Step 2}}.
We then check to see if the block $b_r$ is involved in any \textit{method-body-level} refactorings, such as \textsc{Replace Loop with Pipeline}, \textsc{Invert Condition}, \textsc{Split Conditional}, \textsc{Merge Conditional}, \textsc{Merge Catch}. If that is the case, we report the appropriate refactoring as a change on the block and continue the tracking process with the matching statement $b_p$ in commit $p$ reported in the refactoring instance.
Note that merge-related refactorings will introduce a fork in the evolution history of the tracked block.

In the case none of the aforementioned refactorings were performed involving $b_r$, we obtain the statement mappings returned by RefactoringMiner and check if $b_r$ has been mapped to a statement $b_p$ in the parent commit $p$.
Upon finding a match, we construct the unique identifiers of $b_r$ and $b_p$ and link the two code element nodes in the graph. 
We then check if the contents of the block have remained the same. If the expression and/or body have changed, we report an \textsc{Expression change} and \textsc{Body change}, respectively. If the block is a \texttt{try} block, we separately compare the contents of the \texttt{catch} blocks. We also report \texttt{catch} blocks that do not have a mapping in the parent/child commit with a \textsc{Catch block added/removed} change, respectively. This approach is also adopted for \texttt{finally} blocks that may be present within the \texttt{try} statement. 

If we don’t find a match for $b_r$ in the above process, we can suspect that $b_r$ may have been introduced in method $M_r$.
There are two possible scenarios. Either $b_r$ corresponds to new functionality added in method $M_r$, or $b_r$ has been moved to $M_r$ by inlining a method originally called by $M_p$.
To verify the latter scenario, we check if RefactoringMiner reported an \textsc{Inline Method} with $M_r$ as the target, and $b_r$ has been matched with a block $b_p$ from the inlined method.
In that case, the tracking will continue with block $b_p$ located in the inlined method from commit $p$.
To verify the first scenario, we check through RefactoringMiner’s list of unmapped blocks present in $M_r$ and then check if any of those correspond to $b_r$. If that is the case, we can safely say that the block has been \textit{Introduced} in commit $r$ as part of a newly added functionality of a bug fix.
If $b_r$ is not found within this list, we need to move on to \textbf{\textsc{Step 4}}, where we explore the possibility of $b_r$ belonging to a method whose signature (i.e., parameter type list, method name, return type) changed, or method $M_r$ is introduced as the outcome of a local \textit{intra-file-method} refactoring, such as \textsc{Extract Method}, \textsc{Merge Method}, \textsc{Split Method}.

\header{\textsc{Step 4}. Check if the container method signature changed} 

At this point, we utilize the information extracted by RefactoringMiner in \textbf{\textsc{Step 3}} after its execution on the partial source code models for commits $r$ and $p$.
RefactoringMiner initially matches the method pairs within type declaration containers ($CON_{M_p}$, $CON_{M_r}$) with identical
signatures (i.e., method name and parameter type list), and then compares all combinations of the remaining unmatched methods from
$CON_{M_r}$ with the remaining unmatched methods from $CON_{M_p}$ to find the best matching method pairs with changes in their signatures.
If $M_r$ is found in a matching method pair ($M_p$, $M_r$) with method signature changes (e.g., \textsc{Rename Method}, \textsc{Add Parameter}, \textsc{Change Parameter Type} refactoring), we obtain the statement mappings returned by RefactoringMiner and check if $b_r$ has been mapped to a statement $b_p$ in the parent commit $p$.
If indeed a statement mapping is found for $b_r$, we construct the unique identifiers of $b_r$ and $b_p$, and link the two code element nodes in the graph.
Otherwise, we utilize the \textit{intra-file-method} refactoring information extracted by RefactoringMiner to examine if any of the remaining unmatched methods from $CON_{M_r}$ has been extracted from a pair of matched method pairs (i.e., \textsc{Extract Method} refactoring), or if any subset of the remaining unmatched methods from $CON_{M_p}$ have been merged to a single remaining unmatched method from $CON_{M_r}$ (i.e., \textsc{Merge Method} refactoring), as well as the reverse scenario (i.e., \textsc{Split Method} refactoring).
If $M_r$ is found being involved in any of the aforementioned refactoring scenarios, we obtain again the statement mappings included in the corresponding refactoring instance and check if $b_r$ has been mapped to a statement $b_p$ in the parent commit $p$.
If a statement mapping is found for $b_r$, we construct the unique identifiers of $b_r$ and $b_p$, link the two code element nodes in the graph,
and continue the tracking process with block $b_p$ from commit $p$.

If no matches are found in this step, then we move on to \textbf{\textsc{Step 5}}, where we check if $b_p$ is located in a file other than \texttt{filePath} in which $b_r$ is located, since there is a possibility that the container method $M_r$ has been moved to \texttt{filePath} from another file, or the type declaration $CON_{M_r}$ containing $M_r$ has been renamed or moved to another package.

\header{\textsc{Step 5}. Include additional files in the partial models}

This step is the most computationally expensive step of the tracking process, as we keep the partial source code model for commit $r$ as is, but add all modified and removed files in commit $p$ to $p$’s source code model (i.e., we create the complete source code model for commit $p$) to enable the detection of \textit{inter-file-method} refactorings, such as \textsc{Pull Up}, \textsc{Push Down}, \textsc{Move Method}, as well as \textit{class-level} refactorings, such as \textsc{Move}, \textsc{Rename}, \textsc{Extract}, \textsc{Merge}, \textsc{Split Class}.
To avoid the unnecessary processing of files and
speed-up the tracking process, we exclude from $p$’s source code
model all files with identical contents, and files with only trivial
changes in comments (e.g., license headers) and import declarations~\cite{Non-Essential-Changes}.
Moreover, we support two scenarios in which additional files need to be included
in $r$’s source code model to correctly track $b_r$:
\begin{enumerate}[wide,leftmargin=*,labelwidth=!, labelindent=0pt]
	\item \header{$b_r$ is copied into a new file}
	In some projects, which are libraries with public APIs, we found that developers tend to copy the methods they want to deprecate into a new file, and then declare the original methods or their container class as \texttt{@deprecated}.
	Let us assume that $M_r$ (i.e., the method containing $b_r$) is copied in type declaration $T_r$ in commit $r$ from type declaration $T^\prime_p$ in commit $p$.
	Without additionally including the original type declaration containing the copied method $T^\prime_r$ to $r$'s source code model, then $M_r$ would be detected as \textit{moved} from $T^\prime_p$ to $T_r$, instead of \textit{introduced} in $T_r$ as a new method.
	To address this issue we use a regular expression to check if other modified files in commit $r$ include a \texttt{@deprecated} annotation with a \texttt{@link} to $M_r$'s signature
	(e.g., \texttt{copy} methods copied from \texttt{IOUtils} to \texttt{CopyUtils} in project commons-io~\cite{Commons-IO-1}),
	or a \texttt{@deprecated} annotation with a reference to $T_r$ name (e.g., deprecated classes \texttt{IOUtil} and \texttt{EndianUtil} referring to newly added classes \texttt{IOUtils} and \texttt{EndianUtils}, respectively, in project commons-io~\cite{Commons-IO-2})
	and add them to $r$'s source code model.
	Moreover, we check if other modified files in commit $r$ have the same name as $T_r$, but a different package (e.g., methods copied from deprecated class \texttt{org.apache.commons.lang.NumberUtils} to new class \texttt{org.apache.commons.lang.math.NumberUtils} in project commons-lang~\cite{Commons-Lang}) and add them to $r$'s source code model.
	\item \header{$b_r$ is extracted to a new file}
	In this scenario, developers move some members of an existing class into a new class, and instantiate the new class into the origin class in order to access the moved functionality (i.e., \textsc{Extract Class} refactoring), or extend the origin class in order to inherit the non-moved functionality (i.e., \textsc{Extract Subclass} refactoring).
	Let us assume that $M_r$ (i.e., the method containing $b_r$) is moved in type declaration $T_r$ in commit $r$ from type declaration $T^\prime_p$ in commit $p$.
	Without additionally including the original type declaration containing the moved method $T^\prime_r$ to $r$'s source code model, then $T^\prime_p$ would be detected as
	\textit{renamed} to $T_r$ (if multiple members from $T^\prime_p$ have been moved to $T_r$), instead of $T_r$ being extracted from $T^\prime_p$, and $T^\prime_r$ being matched with $T^\prime_p$.
	To address this issue we use a regular expression to check if other modified files in commit $r$ create an instance of $T_r$
	(e.g., methods moved to extracted class \texttt{SourceFileInfoExtractor} from class \texttt{ProjectResolver} in project javaparser~\cite{JavaParser}),
	or are extended by $T_r$ (e.g., methods pushed down to extracted subclass \texttt{AbstractNestablePropertyAccessor} from origin class \texttt{AbstractPropertyAccessor} in project spring-framework~\cite{Spring-Framework}) and add them to $r$'s source code model.
\end{enumerate}

After including any additional files needed in $r$’s source code model, we go through all pairs of files that belong in both partial source code models for commits $r$ and $p$
and remove all pairs of method declarations that are identical, as these methods cannot serve as candidates for a move.
This optimization of the partial source code models reduces the processing time for RefactoringMiner's execution and also reduces the chances of mismatching $M_r$ with an irrelevant but rather similar method from commit $p$.

\header{\textsc{Step 5a}. Check if the container class is moved/renamed}

After setting up the partial source code models for commits $r$ and $p$, we execute RefactoringMiner again.
First, we check all \textit{class-level} refactorings (e.g., \textsc{Move Class}, \textsc{Rename Class}) to find a pair of type declarations ($CON_{M_p}$, $CON_{M_r}$) involving $CON_{M_r}$.
If such a pair is found, we obtain the corresponding class-level diff object from RefactoringMiner, which includes all pairs of matched methods.
We then check if $M_r$ is included in the matching method pairs.
If so, we obtain the statement mappings returned by RefactoringMiner for the ($M_p$, $M_r$) method pair, and check if $b_r$ has been mapped to a statement $b_p$ in the parent commit $p$.
If indeed a statement mapping is found for $b_r$, we construct the unique identifiers of $b_r$ and $b_p$, and link the two code element nodes in the graph.

\header{\textsc{Step 5b}. Check if $b_r$ is moved to another file}

If there is still no match found for $b_r$, this is an indication that either $b_r$ itself or its method container $M_r$
has been moved to another file through an \textsc{Extract and Move Method} or \textsc{Move Method} refactoring, respectively.
To assert this scenario, we check in all \textit{inter-file-method} refactorings reported by RefactoringMiner if method container $M_r$ is involved.
If so, we obtain the statement mappings included in the corresponding refactoring instance and check if $b_r$ has been mapped to a statement $b_p$ in the parent commit $p$.
If indeed a statement mapping is found for $b_r$, we construct the unique identifiers of $b_r$ and $b_p$, and link the two code element nodes in the graph.

If by the end of \textbf{\textsc{Step 5}} there is still no match found for $b_r$, we report that $b_r$ has been \textit{Introduced} in commit $r$ as part of a newly added method.

Steps 2-5 are iteratively executed until the tracked block is found as \textit{Introduced}, or until we reach the first commit of the project, which means that the tracked block has existed since the beginning of the project.

\subsection{Tracking Algorithm Completeness}
Each step of our algorithm has been designed to address a specific scenario related to the possible \textit{location} and \textit{transformation state} of the tracked block.
Each step incrementally broadens the search scope for the tracked block by computing additional refactoring information and including more relevant files in the analysis.

\begin{enumerate}[wide,leftmargin=*,labelwidth=!, labelindent=0pt]
	\item \textbf{\textsc{Step 2}} covers the scenario where the method containing the block remained \textit{unchanged}. Note that the file containing the method is changed, otherwise it would not be returned by the \texttt{git log} command in \textbf{\textsc{Step 1}}.
	\item \textbf{\textsc{Step 3}} covers the scenario where the method containing the block preserves an \textit{identical signature}, but its \textit{body has changed}.
	There are 4 possible sub-scenarios:
	\begin{enumerate}[leftmargin=*,labelindent=2pt]
		\item The tracked block remained \textit{unchanged}, i.e., the changes affected other parts of the method.
		\item The tracked block had a \textit{change within its own body, or conditional expressions}.
		\item The tracked block has been \textit{migrated} to a different block structure. This sub-scenario includes all 9 \textit{block-to-block} transformations explained in Section~\ref{sec:block-tracking-process}. 
		\item The tracked block has been \textit{refactored} with
		\textit{method-body-level} and \textit{block-specific} refactorings, such as \textsc{Replace Loop with Pipeline}, \textsc{Split Conditional}, \textsc{Merge Conditional}, \textsc{Invert Condition}, \textsc{Merge Catch}, etc., which affect the control structure of the method.
	\end{enumerate}
	\item \textbf{\textsc{Step 4}} covers the scenario where the tracked block belongs to a \textit{method with a different signature} within the same file. There are 2 possible sub-scenarios:
	\begin{enumerate}[leftmargin=*,labelindent=2pt]
		\item The \textit{signature of the method containing the block changed}, due to an addition/deletion of parameter, change of parameter type or return type, change of method name, etc.
		This \textit{method already existed before}.
		\item Some \textit{intra-file-method} refactoring affected the tracked block. For example, the tracked block could have been \textit{extracted} from another method (possibly more than one method in case of duplicated code extraction), or the tracked block could be located in a new method resulting after \textit{merging} two or more previously existing methods, or the tracked block could be located in a new method resulting after \textit{splitting} a previously existing method into two or more methods. In all these scenarios, the tracked block is \textit{located in a method that did not exist before}.
	\end{enumerate}
	\item \textbf{\textsc{Step 5}} covers the scenario where the tracked block is \textit{no longer located in the same file} (i.e., a file with the same file path). It is divided in two sub-steps.
	\begin{enumerate}[leftmargin=*,labelindent=2pt]
		\item \textbf{\textsc{Step 5a}} covers the case where the file containing the block has been \textit{moved/renamed}, or it has been \textit{extracted} from another file, or resulted after \textit{merging} two or more previously existing classes, or resulted after \textit{splitting} a previously existing class into two or more classes.
		In all these cases, the tracked block is located in a \textit{file that did not exist before with the same file path}.
		\item \textbf{\textsc{Step 5b}} covers the case where the block itself has been moved to another \textit{pre-existing file}. This can be done either with an \textit{inter-file-method} refactoring \textit{moving the entire method} containing the block (\textsc{Move Method}, \textsc{Pull Up Method}), or an \textit{inter-file-method} refactoring \textit{moving part of the method} containing the block (\textsc{Extract and Move Method}).
	\end{enumerate}
\end{enumerate}
Collectively, the aforementioned steps cover every possible scenario regarding the location of the tracked block.
Moreover, the aforementioned refactorings detected by RefactoringMiner cover all possible ways a block and its container method can be refactored according to Fowler's catalogues~\cite{Fowler:1999, Fowler:2018}.

\subsection{CodeTracker API and Chrome Browser Extension}
\label{sec:chrome-extension}
\begin{figure}[ht]
	\centering
	\vspace{-2mm}
	\includegraphics[width=\linewidth]{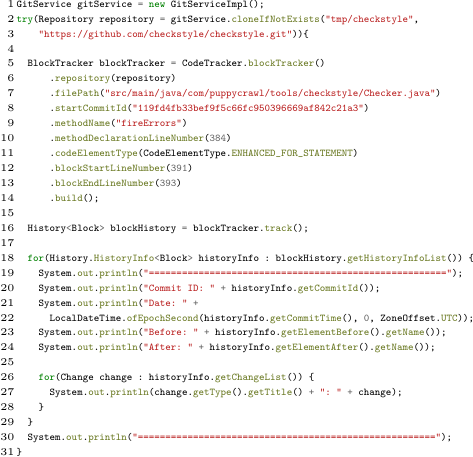}
	\vspace{-7mm}
\caption{Fluent API for block tracking}
\vspace{-3mm}
\label{listing:api}
\end{figure}

CodeTracker~\cite{CodeTrackerOracle} is available in Maven Central Repository~\cite{codetracker-maven} and can be used as a library within Java projects.
It offers a set of fluent APIs for tracking code blocks, methods, variables, and attributes.
Figure~\ref{listing:api} is a code snippet demonstrating the fluent API usage for block tracking.
Lines 1-3 clone the repository \texttt{checkstyle} into the local directory \texttt{tmp/checkstyle}, if the project is not already cloned, and create the object \texttt{repository} as an instance of the JGit library \texttt{Repository} type.
Lines 5-14 create an instance of \texttt{BlockTracker} using the fluent builder pattern.
Line 16 executes the tracking process for the specified \texttt{blockTracker}.
Lines 18-30 iterate over the \texttt{HistoryInfo} elements within \texttt{blockHistory} and print commit and change related information.

\begin{figure*}[h]
    \centering
    \includegraphics[width=\linewidth]{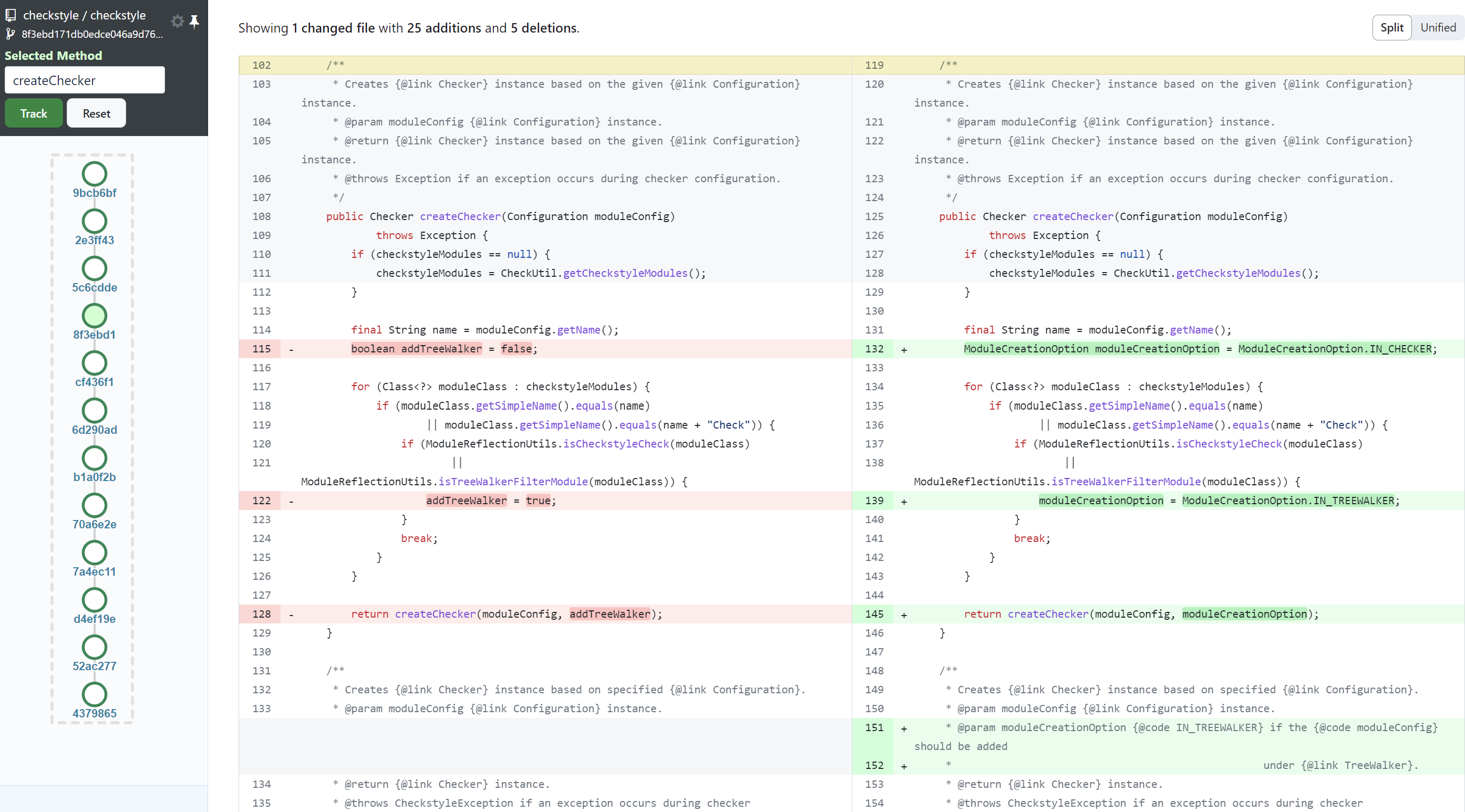}
    \vspace{-5mm}
    \caption{CodeTracker Chrome browser extension visualizing the change history for a selected program element.}
    \vspace{-3mm}
    \label{fig:full-view-extension}
\end{figure*}

Moreover, CodeTracker is available as a Chrome browser extension~\cite{codetracker-visualizer}, which
 integrates with the GitHub web UI to provide a visual overlay with code change history for any code element
present on GitHub.
Figure~\ref{fig:full-view-extension} shows a screenshot visualizing the change history of method \texttt{createChecker} from project \texttt{checkstyle}.
After installing the Chrome extension, the user can obtain the change history for a code element (method, variable, attribute, or code block)
by loading a GitHub commit page or file blob page in the browser
and double-clicking on the desired code element.
When the user selects a code element, we capture the mouse event and obtain the text selected by
the user. This selection should be the name of a method, attribute, variable, or a code block Java keyword (e.g., \texttt{if}, \texttt{for}).
By accessing the DOM, we then pick up the line at which the code element is present and capture
the line number. After this, we move up in the DOM until we reach the file container of the line,
which contains the file path of the class containing the selected code element. Finally, we capture
the commit from the webpage URL, along with the repository name.
All this information is then passed to CodeTracker’s REST API, which is run on a Java Web Server and can
serve CodeTracker’s functionalities over the web.
The REST API endpoint \texttt{GET codeElementType} takes
as input the information provided above, and returns the type of code element being selected, which is then displayed to the user on the side panel along with the name of the code
element, as shown in Figure~\ref{fig:full-view-extension} top-left corner (i.e., \texttt{Selected Method createChecker}). 
This helps provide instant feedback to the user about the validity of her selection with just a click.
When the user makes an invalid selection, e.g., an incomplete method name, an unsupported Java keyword, an
operator, or multiple code elements at once, we gray out the track button using the information obtained
from this API endpoint.
When the user has made a valid selection the “Track” button is enabled.
Clicking on the “Track” button initiates the
tracking process using CodeTracker, via its REST API. Once the tracking data is
obtained from the REST API, we model the JSON response into a graph, which is essential to render a
visual representation of the change history evolution, as displayed on the left-side panel in Figure~\ref{fig:full-view-extension}.

\begin{figure}[h]
    \centering
    \includegraphics[width=\linewidth]{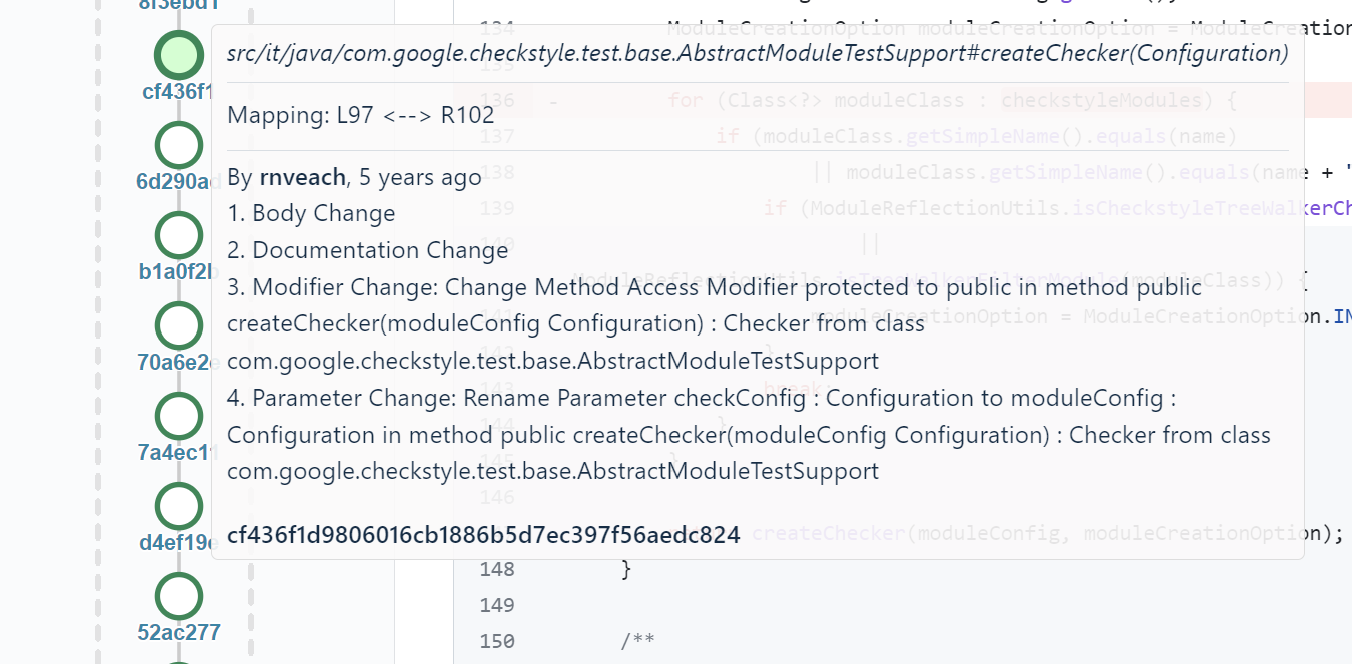}
    \vspace{-7mm}
    \caption{Hovering over a node provides more semantic information about the changes that occurred on the tracked code element.}
    \vspace{-3mm}
    \label{fig:hover}
\end{figure}
Each green node on the left-side panel indicates a commit in which one or more changes occurred on the tracked code
element. The nodes are sorted chronologically starting from the most recent commit (on the top) and ending with the commit in which the tracked code element was introduced (on the bottom).
As seen in Figure~\ref{fig:hover}, when the user hovers their cursor over a node, a tooltip appears with a semantic description of the
change(s) and commit-related information, such as the commit author and time.
If the user clicks on a node, the corresponding GitHub commit webpage will load and the user will be navigated to the exact line of the code
element within this specific commit, by automatically scrolling and expanding hidden parts of the GitHub diff if needed.
This feature allows the user to quickly inspect the changes, and possibly confirm a feature implementation, a bug fix, a bug introduction, or some behavior-preserving changes due to refactorings.

The ease of use was a key element in mind during the design and development of the Chrome browser extension, and our approach emphasizes this aspect.
A user can obtain the entire code change history for a code element with just two clicks (i.e., a double-click to specify the code element of interest and a click on the “Track” button to initiate the tracking process), and can navigate to the exact location of the tracked code element at any commit with one additional click. This streamlined process aims to make the usage of CodeTracker as a code change history generator more efficient.
Moreover, as discussed later on in Section~\ref{sec:oracle-creation}, we used a slightly modified version of the Chrome browser extension to validate and create our block change oracle.

\subsubsection{How can the Chrome extension help answering common developer questions about the evolution of blocks}
\label{sec:chrome-extension-answer-questions}
The questions that follow have been listed as commonly asked questions by professional developers in various surveys~\cite{LaToza:Hard-to-Answer-Questions, Fritz:2010, Ko:2007}:

\noindent
``\textit{Where was this block last changed?}''
The top node on the left-side panel in Figure~\ref{fig:full-view-extension} shows the commit where the selected block was last changed.

\noindent
``\textit{When, how, by whom, and why was this block inserted?}''
The bottom node on the left-side panel in Figure~\ref{fig:full-view-extension} shows the commit where the selected block was initially introduced.
Hovering over the node (Figure~\ref{fig:hover}) shows the developer that first introduced the block, the date the block was introduced, and a more fine-grained description of the block introduction, explaining whether the block was introduced in an already existing method, or as part of a newly added method.
Clicking on the node will load the corresponding commit in the browser navigating the user to inspect how exactly the block was added in its container method.
The \textit{why} could be perhaps answered by the text or linked issues in the commit message, but CodeTracker does not offer any particular help to find why the block was introduced.

\noindent
``\textit{How has this block changed over time?}''
All nodes shown on the left-side panel in Figure~\ref{fig:full-view-extension} represent the change history of the selected block.
Hovering over each node (Figure~\ref{fig:hover}) shows the developer that performed the change, the date the change was committed, and a more fine-grained description of the change(s)
including \textit{block-to-block} migrations and \textit{block-specific} refactorings.
Clicking on a node will load the corresponding commit in the browser navigating the user to inspect how exactly the block changed.
\section{Evaluation}
In our evaluation, we investigate the following research questions:
\begin{enumerate}[wide,label=\bfseries RQ\arabic*., leftmargin=*,labelwidth=!, labelindent=0pt]
	\item What is the accuracy of CodeTracker in block tracking and how does it compare to that of a baseline approach based on the GumTree AST diff tool?
	\item How does the execution time of CodeTracker compare to that of the baseline?
\end{enumerate}

\subsection{Oracle creation}
\label{sec:oracle-creation}
Grund et al.~\cite{Grund:2021} created an oracle with the change history of 200 methods from 20 popular
open-source project repositories.
In particular, they used 100 of these methods (\textit{training set}) to optimize the threshold values used in CodeShovel (i.e., their tool extracting method change history), until they achieved 100\% training accuracy, and the remaining 100 methods (\textit{testing set}) to validate the accuracy of CodeShovel.

Later on, Jodavi and Tsantalis~\cite{Jodavi:2022} re-validated the oracle by Grund et al. and corrected some discrepancies.
More specifically, they found that 18 methods from the
\textit{training set} and 9 methods from the \textit{testing set} were matched with
a method extracted from their body at some point in their change
history. As a result, the Grund et al. oracle includes the change history of the extracted method, instead of the originally tracked method.
Apart from correcting the aforementioned discrepancies, Jodavi and Tsantalis extended the oracle
with the change history of the local variables and parameters declared in these 200 methods (967 variables in the \textit{training set} and 378 variables in the \textit{testing set}).

\begin{figure}[h]
    \centering
    \vspace{-3mm}
    \includegraphics[width=\linewidth]{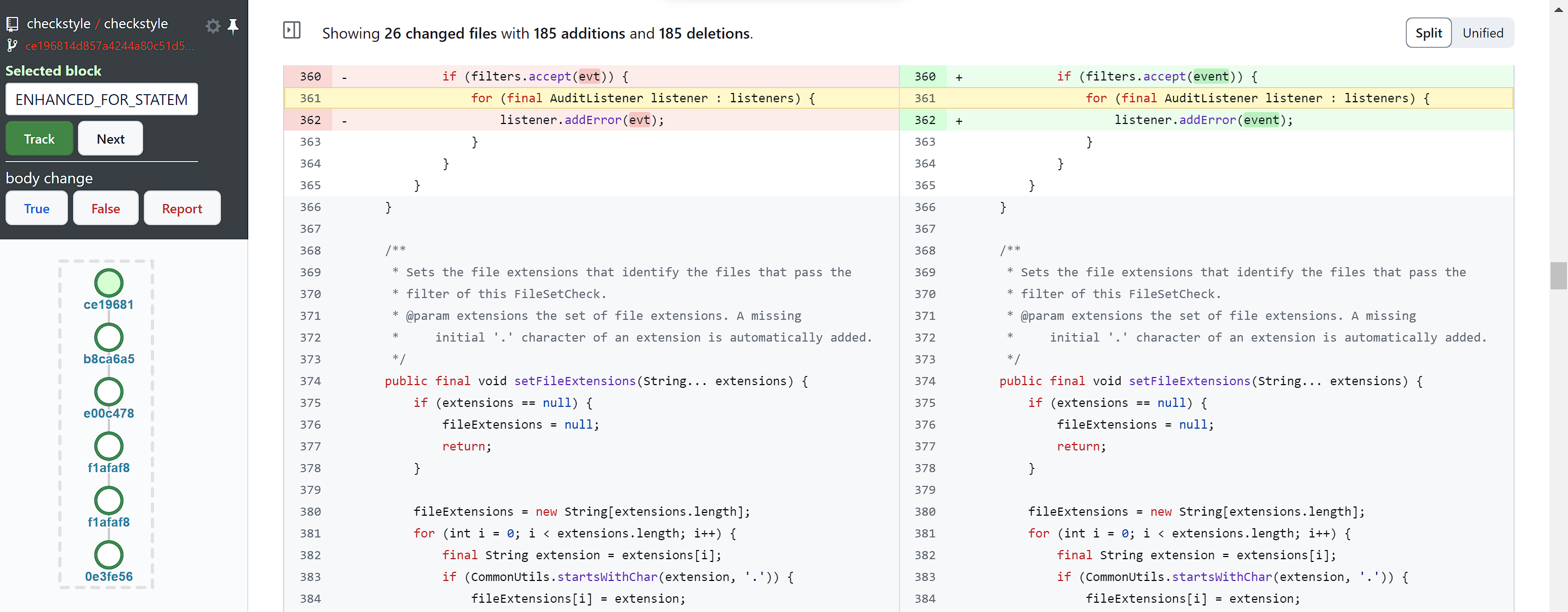}
    \vspace{-7mm}
    \caption{Chrome browser extension used for validating the oracle.}
    \vspace{-1mm}
    \label{fig:oracle-validator}
\end{figure}

In this work, we further extend this oracle with the change history of 1,280 blocks included in these 200 methods (964 blocks in the \textit{training set} and 316 blocks in the \textit{testing set}).
To validate the block changes, we combined two complementary approaches.
First, we leveraged information from the method tracking oracle,
as we know for sure that the commits in which a block changed
are a subset of the commits in which its container method changed.
Second, we slightly modified our extension for the Chrome browser discussed in Section~\ref{sec:chrome-extension} to help us in the change validation process, as shown in Figure~\ref{fig:oracle-validator}.
The human validator (i.e., one of the paper authors) can navigate over the commits by clicking on the nodes appearing on the left sidebar shown in Figure~\ref{fig:oracle-validator}.
The Chrome extension loads the commit selected by the validator and automatically scrolls the webpage to the location of the block on the right side of the diff (this might require expanding hidden parts of the source code diff until the block becomes visible).
The validator can visually inspect the source code diff and confirm or reject the left-side matching block changes reported by CodeTracker.

To consider a pair of blocks (from the left and right side, respectively) as a true match in the ground truth,
the blocks should be sharing some common functionality.
This means that some part of the code within their bodies and conditional expressions should implement the same functionality, even if this code is not structurally similar (i.e., refactored), or is using different/alternative APIs.
On the contrary, if the validated blocks implement a different functionality, then we do not consider this pair as a true match, even if the blocks have the same location in the control structure of their container method.

If the validator confirms a change, this change is persisted in a JSON file. Otherwise, the validator manually edits the JSON file with an entry specifying the left-side block that the currently tracked right-side block should have been matched with.
In such a case, to resume the validation process, CodeTracker is re-executed with the correct left-side block as input starting from the parent of the commit in which the error was found.
The validation process completes when one of the following termination conditions is met:
\begin{enumerate}[leftmargin=*]
	\item The block tracking reaches the commit in which the container method was introduced. 
	This means that the block has existed since the introduction of its container method.
	\item The block is introduced in a commit before reaching the container method introduction commit.
	This means that the block was added as part of some new functionality implemented in the container method.
\end{enumerate}

\begin{figure}[h]
	\centering
	\includegraphics[width=0.5\linewidth]{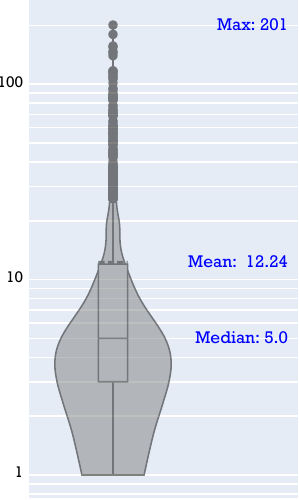}
	\vspace{-4mm}
	\caption{Block size (lines of code) distribution in start commit.}
	\vspace{-3mm}
	\label{fig:size}
\end{figure}

We applied this process for each one of the 1,280 blocks included within the body of these 200 methods (the number of instances per block type is shown in Table~\ref{table:block-types}), and collected a total of 6,093 changes, detailed in Table~\ref{table:block-oracle}, which constitute our block change oracle. 
Figure~\ref{fig:size} shows the size distribution for the 1,280 blocks in their corresponding start commit (the y-axis is in logarithmic scale and the units represent lines of code). The median block size is five lines, while the average size is 12.24 lines. The largest block size is 201 lines, while there are 21 single-line blocks in our dataset.
Overall, the duration of the validation process was three person-months.

\begin{table}[ht]
	\vspace{-4mm}
    \caption{Number of instances per block type included in the oracle}
    \vspace{-4mm}
    \centering
    \begin{tabular}{lr}
        \toprule
        \textbf{Block Type} & \textbf{Number of Instances} \\
        \midrule
        \texttt{if} statement & 929 \\
        \texttt{enhanced-for} statement & 87 \\
        \texttt{try} block & 81 \\
        \texttt{catch} clause & 80 \\
        \texttt{while} statement & 34 \\
        \texttt{synchronized} statement & 23 \\
        \texttt{for} statement & 18 \\
        \texttt{finally} block & 15 \\
        \texttt{switch} statement & 10 \\
        \texttt{do-while} statement & 3 \\
        \midrule
        Total & 1280 \\
        \bottomrule
    \end{tabular}
    \label{table:block-types}
\end{table}

\begin{table}[ht]
	\vspace{-4mm}
	\caption{Number of instances per change type for blocks}
	\vspace{-4mm}
	\centering
	\begin{threeparttable}
		\begin{tabular}{l | r | r |}
			\toprule
			\textbf{Change Type} &
			\textbf{Training set} &
			\textbf{Testing set} \\
			\midrule
                Body Change                  & 3310 & 536  \\
                Introduced                   & 964  & 316  \\
                Expression   Change          & 614  & 120  \\
                Catch   Block Change         & 124  & 40   \\
                Finally   Block Change       & 23   & 2    \\
                Block   Split                & 14   & 0    \\
                Block   Merge\tnote{\textdagger}      & 11   & 3    \\
                Catch   Block Added          & 6    & 6    \\
                Finally   Block Added        & 4    & 2    \\
                Catch   Block Removed        & 2    & 3    \\
                Finally   Block Removed      & 4    & 1    \\
                Replace   Pipeline With Loop & 1    & 0    \\
                Replace   Loop With Pipeline & 1    & 0    \\
			\midrule
			Total                        & 5067 & 1026 \\
			\bottomrule
		\end{tabular}
		\begin{tablenotes}
		\item[\textdagger] not included in the ground truth to ensure a fair comparison with the baseline (Section~\ref{sec:baseline}, last paragraph)
		\end{tablenotes}
	\end{threeparttable}
	\label{table:block-oracle}
	\vspace{-5mm}
\end{table}

\vspace{-2mm}
\subsection{Baselines}
In our previous work~\cite{Jodavi:2022}, we compared the accuracy of CodeTracker in extracting the change history of methods against CodeShovel~\cite{Grund:2021}, as it was the current state-of-the-art at the time.
However, by design, CodeShovel cannot be easily extended to extract the change history of code blocks, as it relies on method body textual similarity to match a pair of methods, and thus the control structure similarity is not taken into consideration.

\vspace{-2mm}
\subsubsection{Block tracking baseline based on GumTree}
\label{sec:baseline}

To create a competitive baseline for comparing our accuracy, we relied on GumTree 3.0~\cite{gumtree, gumtree-hyperparameter-optimization} for providing block mappings, which is the current state-of-the-art Abstract Syntax Tree (AST) diff tool.
GumTree takes as input a pair of source code files, represented as Abstract Syntax Trees, and generates AST node mappings between the two trees, which are then used to compute an edit script that can transform one AST into the other.

The GumTree-based baseline implementation is essentially identical to the way CodeTracker works, but instead of using RefactoringMiner to obtain statement mappings, it uses GumTree.
GumTree can process only a pair of source code files and does not support commit-level analysis to find AST node mappings between different files in the case of a tracked block being moved to another file.
To assist GumTree in this scenario, we adopted the two-round staged
tree-matching approach proposed by Fujimoto et al.~\cite{StagedTreeMatching}.
When a block is moved from file $f_{1_p}$ in the parent commit $p$ to file $f_{2_r}$ in the child commit $r$, in the first stage of matching we execute GumTree with the file pairs ($f_{1_p}$, $f_{1_r}$) and ($f_{2_p}$, $f_{2_r}$) as input.
For the second stage of matching, we execute GumTree with the unmatched nodes from $f_{1_p}$ and the unmatched nodes from $f_{2_r}$ as input.
This approach ensures that only the remaining unmatched nodes can be potentially matched in the second stage, and is applicable only if files $f_1$ and $f_2$ exist in both parent and child commits. 
To avoid applying this approach for all combinations of file $f_{2_r}$ with the parent commit files,
we utilize RefactoringMiner to detect the \textsc{Move Method}, \textsc{Extract and Move Method}, \textsc{Split Class}, or \textsc{Merge Class} refactoring involving the moved block, and provide directly the parent commit file (i.e., $f_{1_p}$) from which the block was moved.

Moreover, Gumtree is designed under the constraint that a given AST node can only belong to one mapping, and thus it cannot match a merged block with each one of the original blocks being merged, but just with one of them.
To ensure a fair comparison with the baseline, 
we did not include in the ground truth the change histories for all forks (i.e., for each one of the blocks being merged), but we included only the change history of the fork corresponding to the original block that has the largest overlap (in source lines of code) with the merged block.
As shown in Table~\ref{table:block-oracle}, our dataset includes 14 instances of block merges, 12 of them are merged \texttt{if} statements and two of them are merged \texttt{catch} blocks.
The source code of the GumTree-based baseline is publicly available~\cite{gumtree-block-tracker}.

\vspace{-2mm}
\subsubsection{Git-log baseline}
As a second baseline, we use the \texttt{git log -L} command, which according to Git documentation~\cite{git-log} it can
``trace the evolution of the line range given by \texttt{<start>,<end>} within the \texttt{<file>}''.
Grund et al.~\cite{Grund:2021} used a similar approach to compare the accuracy of CodeShovel in tracking the change history of method declarations
by providing as arguments the method start and end lines.
In our implementation, we first checkout the repositories to their corresponding start commit, and then execute the \texttt{git log -L} command for each block
with the start and end line of the block in the start commit (this range includes the body of the block), and the file path that the block is located in the start commit.
The command returns a list of commit SHA-1 ids, based on which we can compute the precision and recall of this baseline at commit level.
In some cases, we noticed that \texttt{git log -L} returns commits preceding the block introduction commit in the ground truth, which means that it continued tracing the specified line range beyond the commit where the block was introduced.
To ensure a fair comparison, we decided to exclude such reported commits from the computation of precision and recall, as we consider that \texttt{git log -L} succeeded in tracing the block back to its introduction.
Moreover, after some experimentation with \texttt{git log -L}, we realized that the tracing is getting derailed in commits where the file containing the tracked block has been entirely reformatted with CRLF (i.e, line endings) and/or indentation (i.e., tabs to spaces) changes~\cite{Kim:2006}. To detect such commits, we execute for each reported commit the \texttt{git diff} command with and without the \texttt{--ignore-all-space} and \texttt{--ignore-blank-lines} arguments enabled, and check whether over 95\% of the file lines are reformatted.
In all these cases, we re-run \texttt{git log -L} starting from the reformatting commit by
manually specifying the line range of the tracked block in that commit,
and append
the newly reported commits to the subset of commits returned by the previous execution of the command up to the reformatting commit.
The source code of the \texttt{git log} baseline is publicly available~\cite{gitLog-baseline}.

\subsection{RQ1: Block Tracking Accuracy}
The precision and recall of CodeTracker and GumTree-based baseline were computed at two levels of granularity, namely \textit{commit level} (i.e., finding the commits in which a code block changed), and \textit{change level} (i.e., finding the kinds of changes that occurred in the commits in which a
code block changed).
It should be emphasized that although the results will be presented separately for two datasets, namely \textit{training} and \textit{testing} sets, none of the tools was ``trained'' (i.e., optimized) on the training set.
These datasets are inherited by Grund et al.~\cite{Grund:2021}, who
used 100 methods (\textit{training set}) to optimize the threshold values used in CodeShovel (i.e., their tool extracting method change history), until they achieved 100\% training accuracy, and the remaining 100 methods (\textit{testing set}) to validate the accuracy of CodeShovel.
Both CodeTracker and GumTree-based baseline do not depend on \textit{thresholds} to compare the similarity of program elements, and thus there is no need for training to tune the thresholds.
We preserved the two datasets for the sake of compatibility with previous works that were evaluated on the same datasets~\cite{Grund:2021, Jodavi:2022}.

By design, CodeTracker heavily depends on the statement mapping information generated by RefactoringMiner to match the currently tracked block from the child commit to the corresponding block from the parent commit.
As a result,
the \textit{false positives} (i.e., invalid changes) are due to incorrect statement mappings, while the \textit{false negatives} (i.e., missed changes) are due to RefactoringMiner’s inability to match some pairs of blocks.
\begin{table}[ht]
	\caption{Block tracking precision/recall at commit level}
	\vspace{-4mm}
	\centering
	\setlength\tabcolsep{4px}
	\begin{threeparttable}
		\begin{tabular}{l | l | r r r r r}
			\toprule
			\textbf{Dataset} & \textbf{Tool} & \textbf{TP} & \textbf{FP} & \textbf{FN} & \textbf{Precision} & \textbf{Recall} \\
			\midrule
			\multirow{3}{*}{\textbf{Training}} & GumTree  & \sbar{3873}{220}{840} \\
			                                   & git log -L & \sbar{3922}{1419}{791} \\
			                                   & CodeTracker & \sbar{4701}{8}{12}   \\
			\midrule
			\multirow{3}{*}{\textbf{Testing}} & GumTree  & \sbar{862}{51}{89} \\
						                      & git log -L & \sbar{808}{270}{143} \\
			                                  & CodeTracker & \sbar{950}{4}{1} \\
			\midrule
			\multirow{3}{*}{\textbf{Overall}} & GumTree  & \sbar{4735}{271}{929} \\
			                                  & git log -L & \sbar{4730}{1689}{934} \\
			                                  & CodeTracker & \sbar{5651}{12}{13} \\
			\bottomrule
		\end{tabular}
	\end{threeparttable}
	\label{table:compare-commit}
	\vspace{-5mm}
\end{table}

Based on the results shown in Table \ref{table:compare-commit}, our tool, CodeTracker,
has a consistent performance in both training and testing sets at
commit level, with an overall precision of 99.79\% and recall of 99.77\%.
The GumTree-based baseline has a lower overall precision of 94.55\%, which remains consistent in both training and testing sets. However, there is a considerable difference in recall (8.5\%) for the GumTree baseline between the training set (82.15\%) and the testing set (90.64\%).
This difference can be attributed to two reasons:
\begin{enumerate}[leftmargin=*]
\item The missed block mappings are encountered earlier in the commit change history of the training set compared to the testing set, leading to a longer history of subsequent commits being unprocessed (i.e., false negatives) due to the early miss.
As a matter of fact, the training set has longer commit histories (median: 52.5, average: 59.3 commits) compared to the testing set commit histories (median: 31.5, average: 48 commits). Thus, an early miss in the training set costs more false negatives than an early miss in the testing set.
\item GumTree is a language-agnostic AST diff tool, and thus it can only match nodes of the same AST type. As a result, all cases where a control structure is transformed to another type (e.g., \texttt{for} loop to \texttt{while} loop) are missed mappings for the GumTree-based baseline,
as the AST nodes in these mappings have a different type.
The training set has a total of 16 such control structure transformations, while the testing set has only 6. As a result, there is a larger number of broken change histories in the training set than in the testing set due to control structure transformations, consequently leading to a larger number of unprocessed subsequent commits (i.e., false negatives).
\end{enumerate}

As shown in Table \ref{table:compare-commit}, \texttt{git log -L} has a recall that is quite close to that of the GumTree-based baseline, but has a precision that is considerably lower than the other tools.
By inspecting some results in which \texttt{git log -L} has a large number of false negatives, we found that it prematurely ends the tracing process when the methods in a file get reorganized (i.e., reordered) and the block belongs to a method shown as newly added code in \texttt{git diff}, as happened in project checkstyle for method \texttt{fireErrors()} \cite{Checkstyle-gitLog-FN}. On the other hand, CodeTracker and the GumTree-based baseline are able to correctly match methods within a file, even if they have been reordered.
Furthermore, by inspecting some results in which \texttt{git log -L} has a large number of false positives, we found that it completely derails the tracing process 
when the control flow within the body of a method is restructured (e.g., method \texttt{configure} in hibernate-orm~\cite{hibernate-orm-gitLog-FP}, method \texttt{diff} in jgit~\cite{jgit-gitLog-FP}),
or when the tracked block is moved to a distant location within the file due to \textsc{Extract Method} refactoring (e.g., extracted method \texttt{applyRemovedDiffElement} in javaparser~\cite{javaparser-gitLog-FP}).

\begin{table}[ht]
	\vspace{-2mm}
	\caption{Block tracking precision/recall at change level}
	\vspace{-4mm}
	\centering
	\setlength\tabcolsep{4px}
	\begin{threeparttable}
		\begin{tabular}{l | l | r r r r r}
			\toprule
			\textbf{Dataset} & \textbf{Tool} & \textbf{TP} & \textbf{FP} & \textbf{FN} & \textbf{Precision} & \textbf{Recall} \\
			\midrule
			\multirow{2}{*}{\textbf{Training}} & GumTree  & \sbar{3933}{493}{1135} \\
			                                   & CodeTracker & \sbar{5042}{23}{26}   \\
			\midrule
			\multirow{2}{*}{\textbf{Testing}} & GumTree  & \sbar{865}{114}{161} \\
			                                  & CodeTracker & \sbar{1021}{8}{5} \\
			\midrule
			\multirow{2}{*}{\textbf{Overall}} & GumTree  & \sbar{4798}{607}{1296} \\
			                                  & CodeTracker & \sbar{6063}{31}{31} \\
			\bottomrule
		\end{tabular}
	\end{threeparttable}
	\label{table:compare-change}
\end{table}
Based on the results shown in Table \ref{table:compare-change}, our tool, CodeTracker,
has a consistent performance in both training and testing sets at change level, with an overall precision and recall of 99.5\%.
This is a remarkable accuracy that can mainly attributed to the highly accurate statement mappings generated by RefactoringMiner 3.0.
The accuracy difference between the GumTree-based baseline and CodeTracker is even more intense at the change level, 
as CodeTracker has +11\% in precision and +21\% in recall compared to the baseline.
Next, we will discuss a few indicative false positives and false negatives from both tools to understand better their weaknesses.

\begin{figure*}
    \centering
    \includegraphics[width=\linewidth]{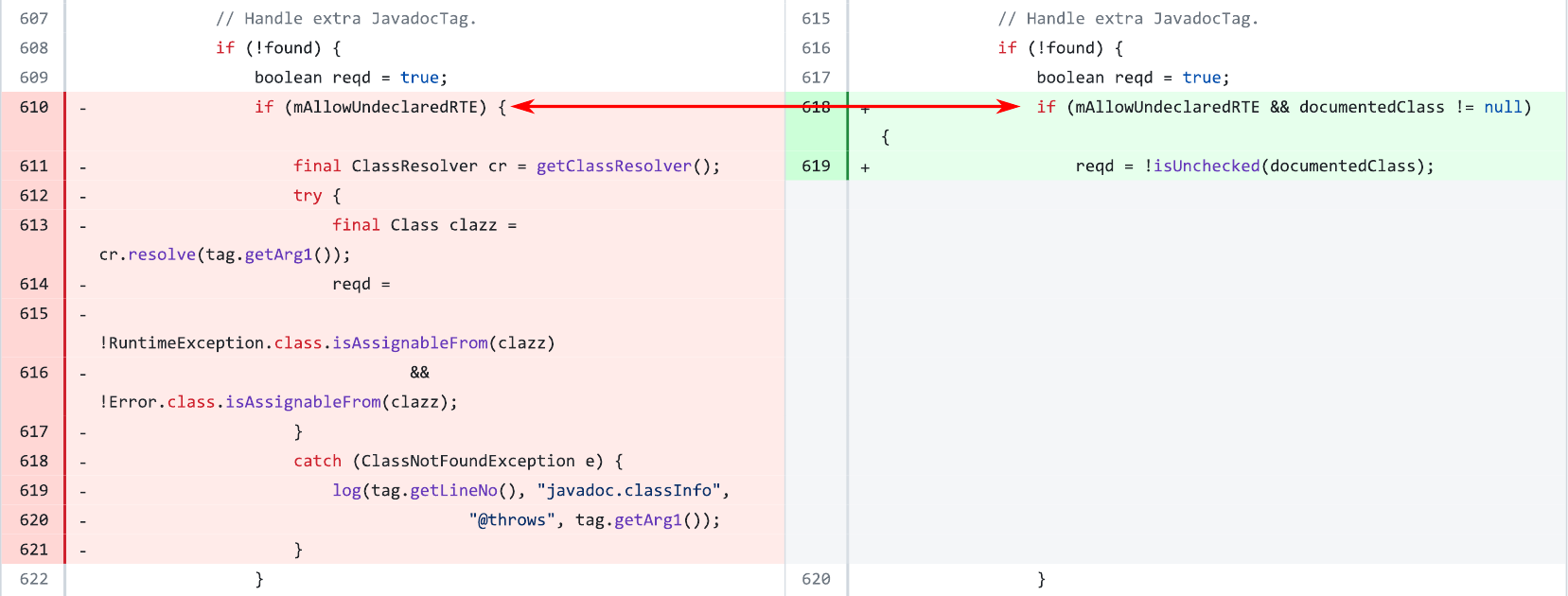}
    \vspace{-7mm}
    \caption{Missed mapping by CodeTracker between L610-622 and R618-620 in commit \cite{Checkstyle-Block-FN}}
    \vspace{-2mm}   
    \label{fig:fn-checkstyle}
\end{figure*}

Figure~\ref{fig:fn-checkstyle} shows a false negative case reported by CodeTracker in project checkstyle~\cite{Checkstyle-Block-FN}.
In this commit, the entire body of block \texttt{if(mAllowUndeclaredRTE)} ranging between lines L611-621 has been deleted and replaced with a single method call \texttt{reqd = !isUnchecked(documentedClass);} in line R619.
RefactoringMiner requires at least one matched pair of statements within the bodies of two blocks (i.e., child statements) in order to match the parent blocks.
In this case, it was not able to establish any child statement mappings, and thus failed to match the parent \texttt{if} blocks L610-622 $\mapsto$ R618-620. Although, it supports the scenario of having the entire body of a block being extracted/inlined to/from a method, in this case \texttt{isUnchecked()} in an already existing method inherited from superclass \texttt{AbstractTypeAwareCheck}.
Moreover, RefactoringMiner can exceptionally match blocks without any child statement mappings if they have identical expressions, but in this case the condition \texttt{mAllowUndeclaredRTE} changed to \texttt{mAllowUndeclaredRTE \&\& documentedClass != null}.
\findings{Insight \#1}{
CodeTracker is unable to match two blocks
when 3 conditions hold at the same time:
\begin{enumerate}[leftmargin=*]
	\item There are zero pairs of nested statements matched within the block bodies
	\item The blocks have non-identical conditional expressions
	\item RefactoringMiner cannot establish that the statements within the body of one block have been extracted to or inlined from another method (i.e., there is an unmatched statement within the other block body calling an extracted or inlined method).
\end{enumerate}
}

\begin{figure*}
    \centering
    \includegraphics[width=\linewidth]{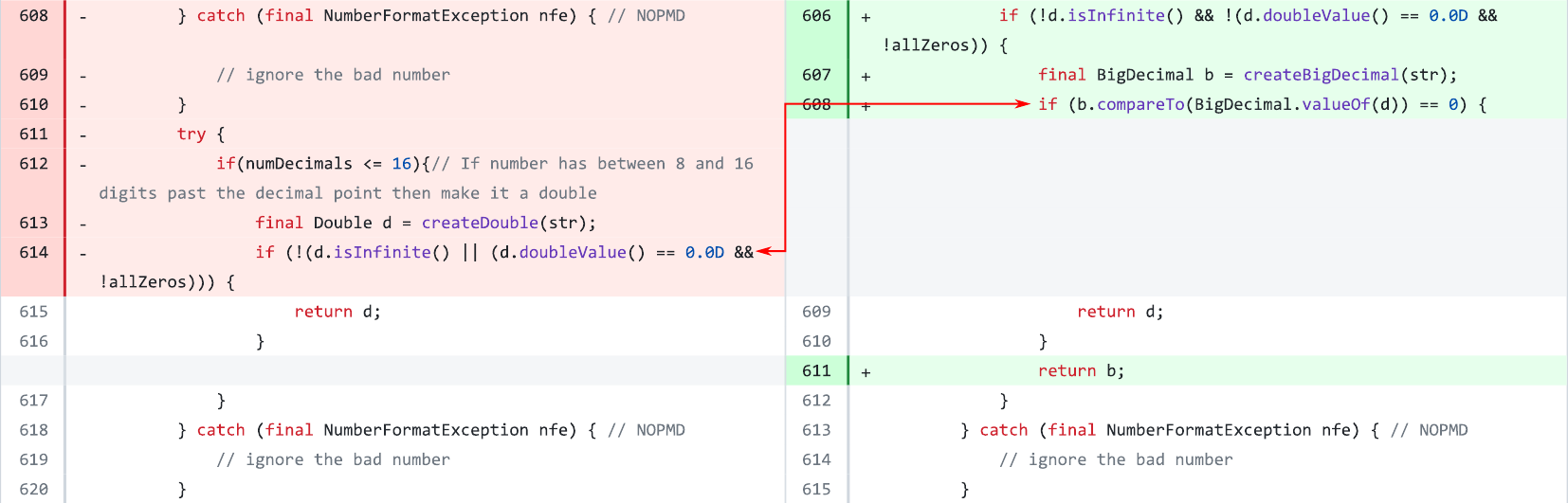}
    \vspace{-7mm}
    \caption{False mapping reported by CodeTracker between L614-616 and R608-610 in commit \cite{Commons-Lang-Block-FP}}
    \vspace{-4mm}   
    \label{fig:fp-commons-lang}
\end{figure*}

Figure~\ref{fig:fp-commons-lang} shows a false positive case reported by CodeTracker in project commons-lang~\cite{Commons-Lang-Block-FP}.
In this commit, there is some control re-structuring that eliminates the \texttt{if} statement ranging between L612-617 and introduces a new \texttt{if} statement ranging between R608-610. At the same time, the \texttt{if} statement ranging between L614-616 is moved to a shallower nesting level ranging between R606-612.
RefactoringMiner finds two candidate mappings for L614-616, namely L614-616 $\mapsto$ R608-610 and L614-616 $\mapsto$ R606-612.
When there are multiple candidate mappings RefactoringMiner uses some ranking criteria to select the best one.
The mapping L614-616 $\mapsto$ R608-610 has two \texttt{if} statements with an identical body (i.e., \texttt{return d;}), and thus its \textit{child match ratio} is perfect (i.e., equal to 1).
On the other hand, the mapping L614-616 $\mapsto$ R606-612 has a \textit{child match ratio} equal to 0.25, because only one out of four child statements in the body of R606-612 is matched.
Therefore, mapping L614-616 $\mapsto$ R608-610 is ranked higher than the other one.
However, the correct mapping is L614-616 $\mapsto$ R606-612, because the expressions of the corresponding \texttt{if} statements are equivalent by simplification (i.e., in R606-612 the \texttt{!} operator and the outermost parenthesis are eliminated).
\findings{Insight \#2}{
	CodeTracker will mismatch two blocks when 3 conditions hold at the same time:
\begin{enumerate}[leftmargin=*]
	\item The control structure of the container method has been restructured
	\item The correct block mapping has a lower \textit{child match ratio} than the mismatched block mapping
	\item The correct block mapping has non-identical conditional expressions.
\end{enumerate}
When there are multiple candidate matches for a block and all of them have non-identical conditional expressions, RefactoringMiner ranks the candidates based on their \textit{child match ratio}.
}

\begin{figure*}
    \centering
    \includegraphics[width=\linewidth]{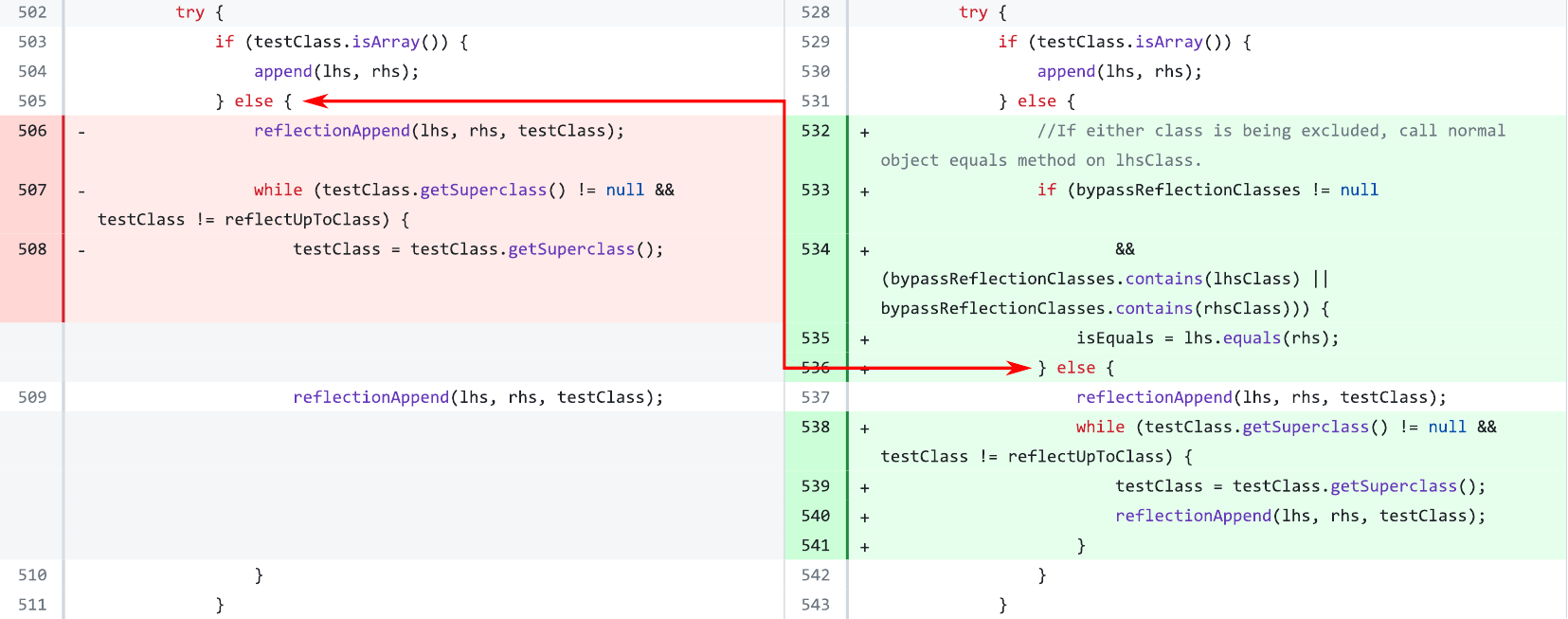}
    \vspace{-7mm}
    \caption{False mapping reported by the GumTree-based baseline between L505-511 and R536-542 in commit \cite{Commons-Lang-Block-GT}}
    \vspace{-4mm}   
    \label{fig:gt-commons-lang}
\end{figure*}

Figure~\ref{fig:gt-commons-lang} shows a false positive and at the same time false negative case reported by the GumTree-based baseline in project commons-lang~\cite{Commons-Lang-Block-GT}.
In this commit, there is some control re-structuring that moves the statements ranging between L506-510 within the \texttt{else} branch (in lines R537-541) of a newly added \texttt{if} statement ranging between R533-542.
GumTree generates mapping L505-511 $\mapsto$ R536-542, instead of L505-511 $\mapsto$ R531-543, which is the correct one.
The reason behind this mistake can be attributed to the way GumTree matches abstract syntax trees in two phases.
In the first phase, GumTree aims to find the largest identical subtrees in a top-down fashion.
In the second phase, GumTree aims to match in a bottom-up fashion the trees that are not matched previously, but at least half of
their children are matched.
Since the \texttt{else} blocks L505-511 and R536-542 are identical subtrees, GumTree matches them in the first top-down phase, without checking whether their parent \texttt{if} statements (i.e., L503-505 and R533-536) are similar enough.
GumTree is designed as a language-agnostic diff tool that matches AST nodes of the same AST type, regardless of their context and semantic role in the program.
However, \texttt{else} branches should not be treated as regular blocks, as they cannot exist in a program without being attached to a parent \texttt{if} statement.
As a result, GumTree should have some language-specific exceptions for matching AST nodes with special characteristics, such as the \texttt{else} branches.
\findings{RQ1 finding}{CodeTracker exhibits an overall precision and recall of 99.5\%.
Compared to the GumTree-based baseline, CodeTracker has +11\% in precision and +21\% in recall.}

\subsection{RQ2: Execution Time}
\begin{figure*}[h]
    \centering
    \includegraphics[width=\linewidth]{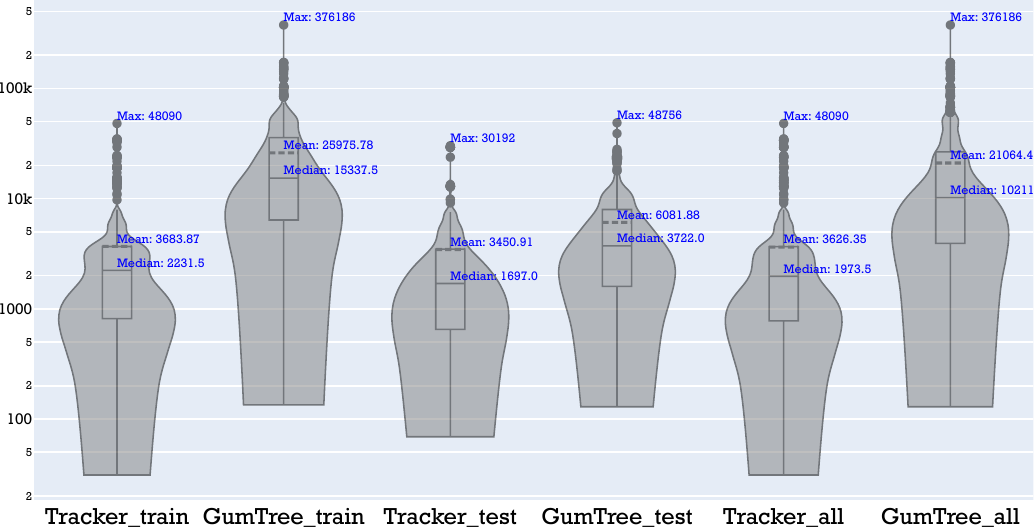}
    \vspace{-6mm}
    \caption{Block change history extraction time in milliseconds.}
    \vspace{-3mm}
    \label{fig:runtimes}
\end{figure*}

Figure \ref{fig:runtimes} shows the execution time of CodeTracker and GumTree-based baseline
for tracking the entire change history of each code block in the training
and testing sets, respectively (the y-axis is in logarithmic scale and
the units are in milliseconds). Each tool was executed separately
on the same machine with the following specifications:
AMD Ryzen 7 5800H CPU @ 3.20GHz × 8, 16 GB 3200 MHz DDR4, 512 GB PCIe SSD, Windows 11 Home operating system, and Java 11.0.15 x64 with a maximum of 8GB Java heap memory (i.e., \texttt{-Xmx8g}). All 20 project repositories used in the oracle were locally cloned before running the tools.
For each tool, we recorded the total time
taken for tracking a code block in its entire commit change history,
including the time taken for parsing the source code files and detecting the changes that took place on the tracked block in each commit, using the
\texttt{System.nanoTime} Java method.

As explained in Section~\ref{sec:baseline}, the GumTree-based baseline relies on RefactoringMiner to detect possible \textsc{Move Method}, \textsc{Extract and Move Method}, \textsc{Split Class}, or \textsc{Merge Class} refactoring involving the tracked block.
To ensure a fair comparison with respect to execution time, we decided to avoid the overhead of RefactoringMiner's execution by persisting into a JSON file the commits in which such refactorings took place and querying this cache during the execution of the GumTree-based baseline to retrieve the origin file path of moved code blocks in constant time.

As we can observe from Figure \ref{fig:runtimes}, CodeTracker has a stable execution time in both training and testing sets (being slightly faster in the testing set), with an overall median execution time of 2 seconds and an average execution time of 3.6 seconds, which makes it suitable for real-time usage when a developer wants to inspect on-demand the change history of a block (assuming the developer is working on a locally cloned repository).

The GumTree-based baseline is significantly slower than CodeTracker (5 times slower on median and 6 times slower on average) with an overall median execution time of 10 seconds and an average execution time of 21 seconds.
Moreover, we can observe a significant difference in the execution time between the training and testing sets.
The GumTree-based baseline is around 4 times slower on both median and average in the training set compared to the testing set.
Table~\ref{table:step-percentage} explains this difference in execution time.
The second column shows the total number of commits processed in the training and testing sets, respectively.
The third column shows the number and percentage of commits in which the container method did not change, and thus there is no need to execute GumTree or RefactoringMiner (this corresponds to \textsc{\textbf{Step 2}} of CodeTracker's approach).
The fourth column shows the number and percentage of commits in which the container method changed, and thus GumTree and RefactoringMiner need to be executed on a pair of files (this corresponds to \textsc{\textbf{Step 3}} and \textsc{\textbf{Step 4}} of CodeTracker's approach).
Finally, the fifth column shows the number and percentage of commits in which the tracked block is moved to another file, and thus GumTree and RefactoringMiner need to include additional files to perform staged tree matching and move detection, respectively (this corresponds to \textsc{\textbf{Step 5}} of CodeTracker's approach).

\begin{table}[ht]
	\vspace{-2mm}
	\caption{Percentage of commits processed in each step of the tracking process.}
	\vspace{-4mm}
	\centering
	\setlength\tabcolsep{3px}
	\begin{threeparttable}
		\begin{tabular}{l | r r r r}
			\toprule
			\textbf{Dataset} & \textbf{\#Commits} & \textbf{No change} & \textbf{Change} & \textbf{Move} \\
			\midrule
			\textbf{Training} & 61,495 &  39,022 (63.45\%) & 21,282 (34.61\%) & 1191 (1.94\%) \\
			\midrule
			\textbf{Testing}  & 16,104 &  13,420 (83.33\%) & 2,302 (14.29\%)  &  382 (2.37\%) \\
			\bottomrule
		\end{tabular}
	\end{threeparttable}
	\label{table:step-percentage}
\end{table}

As we can observe from Table~\ref{table:step-percentage}, the testing set has a larger percentage of commits (83.3\%) that do not require the execution of GumTree or RefactoringMiner to match the tracked block in the parent commit compared to the training set (63.5\%).
On the other hand, the training set has a larger percentage of commits (36.6\%) that require the execution of GumTree or RefactoringMiner to match the tracked block in the parent commit compared to the training set (16.7\%).
This explains the reason why the GumTree-based baseline is slower in the training set compared to the testing set.
The main reason CodeTracker does not exhibit such a big difference in the execution time between the training and testing sets is because it handles the scenario in which the container method changed in two steps, namely \textsc{\textbf{Step 3}} and \textsc{\textbf{Step 4}}.
\textsc{\textbf{Step 3}} has less computation cost than \textsc{\textbf{Step 4}}, as \textsc{\textbf{Step 3}} computes statement mappings just between a single pair of methods, while \textsc{\textbf{Step 4}} computes statement mappings between multiple method pair combinations within a file to find \textit{intra-file-method} refactorings.
On the other hand, the GumTree-based baseline handles the scenario in which the container method changed by computing the AST diff on a pair of whole files, which has a higher computation cost than simply computing the AST diff on a pair of method declarations.
This implementation choice for the GumTree-based baseline was inevitable, as GumTree does not utilize language-specific information to recognize methods with common signatures and \textit{intra-file-method} refactorings that would allow to narrow down the scope of the matching process.

\begin{figure}[h]
	\centering
	\vspace{-3mm}
	\includegraphics[width=\linewidth]{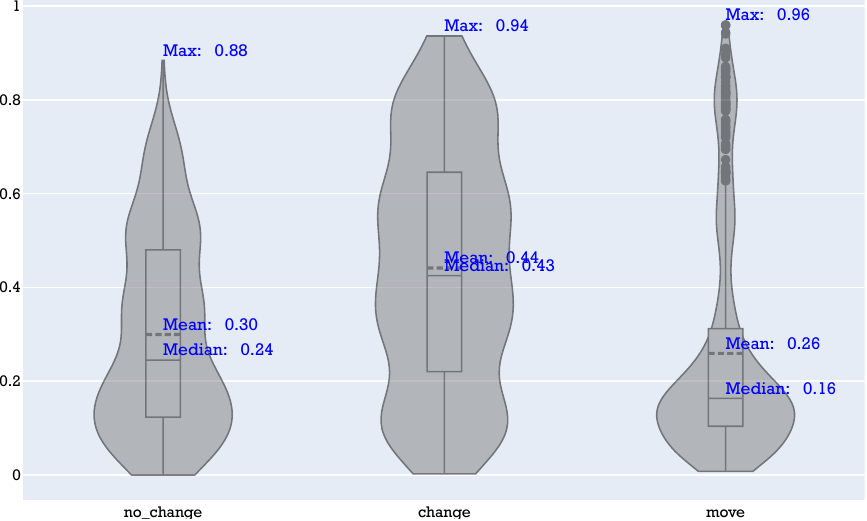}
	\vspace{-7mm}
	\caption{Percentage of CodeTracker's execution time spent on ``No change'', ``Change'' and ``Move'' commits.}
	\vspace{-1mm}
	\label{fig:runtime-percentage-with-moves}
\end{figure}
To give more insights about the performance of CodeTracker, we computed for each of the 1,280 blocks in our dataset the percentage of the total execution time spent on commits in which the container method did not change (``No change'' column in Table~\ref{table:step-percentage}), commits in which the container method changed (``Change'' column in Table~\ref{table:step-percentage}), and commits in which the tracked block is moved to another file (``Move'' column in Table~\ref{table:step-percentage}).
Figure~\ref{fig:runtime-percentage-with-moves} shows the distribution of execution time percentages for 699 out of 1,280 blocks that have at least one commit in their change history where they have been moved to another file.
On average, 30\% of the total execution time is spent on ``No change'' commits, 44\% on ``Change'' commits and 26\% on ``Move'' commits.
Although ``No change'' commits constitute 67\% of the total commits, only 30\% of the total execution time is spent on them.
On the other hand, ``Move'' commits constitute only 2\% of the total commits, but 26\% of the total execution time is spend on them.
\begin{figure}[h]
	\centering
	\vspace{-3mm}
	\includegraphics[width=\linewidth]{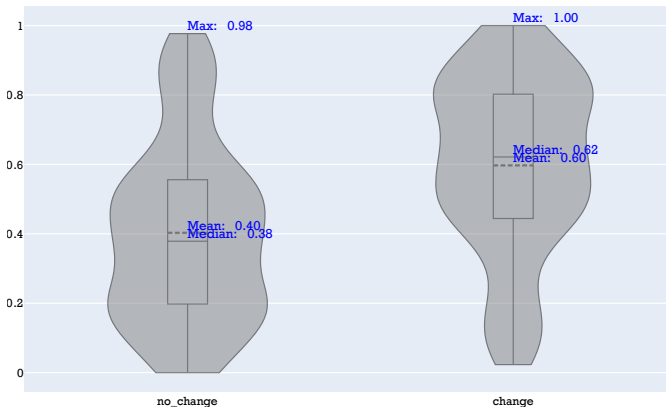}
	\vspace{-7mm}
	\caption{Percentage of CodeTracker's execution time spent on ``No change'' and ``Change'' commits.}
	\vspace{-1mm}
	\label{fig:runtime-percentage-without-moves}
\end{figure}

Figure~\ref{fig:runtime-percentage-without-moves} shows the distribution of execution time percentages for 581 out of 1,280 blocks that have no commit in their change history where they have been moved to another file.
On average, 40\% of the total execution time is spent on ``No change'' commits and 60\% on ``Change'' commits.
\findings{RQ2 finding}{CodeTracker can retrieve the complete change history for a given block within 2 seconds on median and 3.6 seconds on average.
The achieved execution time can
warrant applications in both research (e.g., large-scale MSR and
software evolution studies) and practice (e.g., blame-like tracking of block change history within the context of maintenance and program comprehension tasks).
The GumTree-based baseline was considerably slower in the training set, which had a larger percentage of commits where the method containing the tracked block changed.
}

\subsection{Limitations and Threats to Validity}
\header{Language specificity}
CodeTracker depends on RefactoringMiner 3.0~\cite{Tsantalis:TSE:2020:RefactoringMiner2.0, Alikhanifard:TOSEM:2024:RefactoringMiner3.0} for the detection of refactorings and changes on the tracked program element,
which limits its applicability to Java programs.
Recently, there have been efforts to extend RefactoringMiner for supporting other programming languages, e.g., Python~\cite{PyRef:SCAM:2021,PyRef} and Kotlin~\cite{kurbatova2021refactorinsight,KotlinRMiner}.
Assuming RefactoringMiner supports more programming languages in the future, then
extending CodeTracker to support these languages would require adjusting the program element signature definitions and the regular expressions used in \textsc{\textbf{Step 5}} to the characteristics and structure of these particular languages.
Any language-specific block transformations, such as replacing a \texttt{when} expression with an \texttt{if-else-if} statement in Kotlin, should be supported in the RefactoringMiner core algorithm.
However, we should make clear that making RefactoringMiner support more programming languages is not an easy task.
Although the core statement mapping algorithm is based on string replacements and is not language-specific, 
the algorithm that determines how program element declarations (i.e., method, field declarations) are getting matched is language-specific and depends heavily on the structure and characteristics of Java programs.

\header{Internal validity}
The main threat to internal validity is related to the construction of the oracle used for evaluating precision and recall.
To mitigate this threat we relied on an existing oracle, which was originally constructed by Grund et al.~\cite{Grund:2021} and included the change history of 200 methods, and was later extended by Jodavi and Tsantalis~\cite{Jodavi:2022} by including the change history of 1,345 variables declared within these 200 methods.
Based on this reliable oracle, which was validated independently by two different research groups,
we constructed the change history of 1,280 code blocks declared in the body of these 200 methods
following a semi-automated approach, as explained in Section~\ref{sec:oracle-creation}, and manually inspecting all change instances with the help of our Chrome browser extension.
The overall time dedicated to manually inspecting and validating the change history of 1,280 code blocks was approximately three person-months.
\header{External validity}
Our experiments were conducted on a relatively small dataset including 200 methods from 20 different open-source projects (i.e., 10 methods from each project), which might affected the generalizability of our findings.
However, we decided to design our evaluation experiments on this dataset, as we were already familiar with these methods from our previous work~\cite{Jodavi:2022}, and thus this prior knowledge would speed up the validation process.
\header{Verifiability}
We make the source code of CodeTracker and our extended oracle publicly available~\cite{CodeTrackerOracle} to enable the replication of our experiments and facilitate future research on source code tracking techniques. Moreover, CodeTracker is also available as a Maven library~\cite{codetracker-maven} and as a Chrome browser extension~\cite{codetracker-visualizer} to make easier its usage by researchers and practitioners.
\section{Related Work}
\label{sec:related-work}
\subsection{Line and Statement Tracking}
Canfora et al.~\cite{10.1109/MSR.2007.14, 10.1109/MS.2009.26}
explain that the main problem with the CVS/SVN diff command is
that it cannot detect semantical changes, moves, splits, and
merges of line ranges.
Therefore, they use the output of a CVS/SVN diff command on a pair of files and overcome its limitation by iterating two steps.
The first step compares
ranges of deleted source code lines with ranges of added source code lines, known as \textit{diff hunks}, by computing their \textit{cosine similarity}, i.e., the cosine of the angle between two weighted term
vectors extracted from the deletion and addition line sets.
Ranges with a
similarity greater than a given threshold are assumed to
be in \textit{change relation}.
The second step (\textit{change relation thinning}) further reduces the \textit{change relation}, with the aim of improving the
precision, by computing line-by-line differences using the \textit{Levenshtein} edit distance.
This approach (Ldiff) permits the detection of line
range moves or composition of moves and changes also between different files, otherwise not detectable by using CVS/SVN diff.

Reiss~\cite{10.1145/1368088.1368091} compared 18 different methods for tracking source
locations as their underlying files evolve, and found that
the best-performing method is W\_BESTI\_LINE, which compares lines
using whitespace-insensitive normalized \textit{Levenshtein} edit distance and four lines of context surrounding each line, with a success rate of over 97\%.

Apiwattanapong et al.~\cite{10.1007/s10515-006-0002-0}
defined a new graph representation (\textit{enhanced control-flow graph}) and a differencing algorithm that identifies and classifies changes at the statement
level between two versions of a program. 
The enhanced CFG representation is able to model behaviors caused by object-oriented features in the program, such as dynamic binding, variable and object types, exception handling, synchronization, and reflection.
The algorithm consists of five steps.
First, it matches classes, interfaces, and methods in the two versions. 
Second, it builds enhanced CFGs for all matched methods in the original and modified versions of the program. 
Third, it reduces all graphs to a series of nodes and single-entry, single-exit subgraphs called hammocks. 
Fourth, it compares, for each method in the original version and for the corresponding method in the modified version, the reduced graphs, to identify corresponding hammocks. 
Finally, it recursively expands and compares the corresponding hammocks.
Their hammock matching algorithm is based on Laski and Szermer’s algorithm~\cite{Laski:1992} for transforming two graphs into their
respective isomorphic graphs and takes as input a threshold for deciding whether two hammocks are similar enough to be considered a match.

Spacco and Williams~\cite{10.1109/SCAM.2009.19} proposed
\textit{statement mapping}, a hybrid technique that combines the line-based and structural approaches.
Statement mapping exploits the abstract syntax trees of source
files to break up code into a series of import statements, class
declarations, field declarations, static initializers, and methods,
each of which is made up of a collection of statements.
Statement Mapping effectively ignores any changes to non-functional aspects of code, such as whitespace, curly braces, and comments.
It is also resilient to statement reformatting, i.e., when a programmer breaks a statement across multiple lines.
The algorithm for mapping methods first finds method pairs from the left and right versions of the file having a matching signature and transforms each method into a canonical form.
Each version of the method is represented as a series of statements, which are in turn represented by a series of tokens.
The lists of statements are then compared using the DiffJ algorithm~\cite{diffj}, which generates hunks of differences.
Each diff hunk is treated as a bipartite graph, i.e., each statement
is a node on the appropriate side of the graph and the nodes are connected to each node on the other side of the graph with a weighted edge.
The weight is a similarity metric between the endpoint statements, computed as one of three following metrics:
\textit{normalized Levenshtein edit distance}, \textit{token-based normalized Levenshtein edit distance}, and the minimum of the two.

Servant and Jones~\cite{10.1145/2393596.2393646} developed a \textit{history-slicing} framework, named Chronos.
To build a history graph, Chronos initially utilizes the SCM system’s diff functionality to
determine the added, deleted, and changed individual lines and form \textit{diff hunks}.
In the second phase, it utilizes the Hungarian method for the assignment problem, coupled with Levenshtein distance, to compute an optimal line-to-line mapping within the diff hunks.
The user of Chronos can specify a slicing criterion by opening any revision of any file and selecting any set of lines (contiguous or fragmented).
Next, the history slicer traverses the history graph from the most recent revision of
each line in the slicing criterion, and traces their evolution going backward in time, recording the revisions that contain changes.
The computed history slice is visualized in a zoom-able canvas that depicts all snapshots
for all lines in the slicing criterion, with mappings between them.
In addition, timelines are presented to show proportionally, in time, when changes were made.

\header{Limitations}{
A major limitation of all aforementioned works is that they rely on similarity thresholds to match statements or lines.
This makes them susceptible to overlapping refactorings, such as \textsc{Rename Variable} and \textsc{Extract/Inline Variable} that lower the textual similarity of the statements and even change the original number of statements.
On the other hand, CodeTracker relies on RefactoringMiner, which does not use any similarity threshold to match statements.
Instead, it performs syntactically valid replacements of AST nodes within the statements, until the statements become textually identical. These replacements are then used to infer overlapping edit operations, such as variable renaming, type generalization, and parameter merging.
Moreover, most of the tools rely on the SCM system’s diff to obtain \textit{change hunks}, which are not accurate when code is moved to a distant location, or when code is merged/split. On the other hand, CodeTracker does not rely on any diff information from the SCM, but only uses the \texttt{git log} command to retrieve the commits in which a file changed.
}

\subsection{Program Element Tracking}
Along with the line of work that focuses on tracking specific lines of code or statements, there is a line of work that addresses the same problem at the program element level (i.e., tracking method, attribute, type declarations).

CodeShovel~\cite{Grund:2021}, is the most accurate tool for uncovering Java method histories to date, as it produces complete and accurate commit change histories for 90\% of methods, including 97\% of all method changes.
CodeShovel is partially \textit{refactoring-aware}.
It supports the tracking of methods with changes in their signature (e.g., method rename, parameter addition/deletion), methods whose parent file has been moved/renamed, and methods moved to another file.
However, Jodavi and Tsantalis~\cite{Jodavi:2022} have shown that it fails to track properly methods from which a significant part of their body has been extracted to new methods, as it uses a 75\%  body similarity threshold to match modified methods, and thus erroneously matches the original method with the extracted one.
The same limitation holds when methods with a relatively large body are inlined to the tracked methods.

FinerGit~\cite{FinerGit} and Historage~\cite{Historage} create a finer-grained Git repository, in which each Java method exists in its own file, and
take advantage of
Git mechanisms to track changes in each individual
method’s corresponding file.
FinerGit improves on the limited capability of Historage to track renamed or moved methods, especially for small methods, by formatting each file to include a single token from the corresponding method in each line. This formatting makes Git's line-based similarity computation mechanism more robust in matching small methods, which have been renamed or moved.
Pre-processing an entire repository to place each method in its own file, is computationally expensive and requires additional hard disk space, which can be prohibitive, especially for large repositories with many files and a long commit history.
As a matter of fact, Grund et al.~\cite{Grund:2021} found that FinerGit ran out of memory or did not finish pre-processing within
15 minutes for the four largest repositories in their validation data set.
Moreover, this pre-processing cost did not contribute to an accuracy improvement, as the recall of FinerGit was 65\% compared to 90\% of CodeShovel~\cite{Grund:2021}.

Kim et al.~\cite{Kim:WCRE:2005} proposed an approach to identify function mappings across revisions even when a function's name changes.
Their approach considers the similarity of the following factors: function name, incoming and outgoing calls, signature, function body text diff, complexity metrics, and the results of two clone detection tools (CCFinder and MOSS).
The computation of text diff and the execution of multiple clone detection tools may have a considerable cost, especially when there are many combinations of deleted and newly added functions to be compared.

\header{Limitations} A common limitation of all aforementioned tools is that they are designed to support only the tracking of methods, and cannot be extended to support the tracking of other program elements, such as variables and attributes, whose evolution is also interesting for the developers.
Several studies have shown that developers frequently refactor variables and attributes, which makes their tracking in the commit history challenging.
Negara et al.~\cite{Negara:ECOOP:2013} found that \textsc{Rename Local Variable} and \textsc{Rename Field} are among the most popular refactorings applied by developers.
Negara et al.~\cite{Negara:ICSE:2014} surveyed 420 developers, who ranked \textsc{Change Field Type} as the most relevant and applicable transformation that they perform.
Ketkar et al.~\cite{Ketkar:FSE:2020} found that developers who changed the type of a variable or attribute also renamed it in 55\% of the examined instances.

Godfrey and Zou~\cite{Godfrey:TSE:2005} implemented a tool, named Beagle, that can detect structural changes like rename, move, split, and merge at function, file, and subsystem levels.
They rely on \textit{origin analysis} to decide if a program entity is renamed or moved and a function call analysis to discover merges and splits of program entities.
Although Beagle supports the tracking of program elements at different levels of granularity (i.e., function, file, subsystem), it requires as input two complete versions of a software system in order to extract static relations between program entities (e.g., function calls), and calculate various metrics. This makes Beagle impractical for program element tracking at the commit level.

Steidl et al.~\cite{Steidl:MSR:2014} proposed an incremental origin analysis that applies some heuristics to find moved, renamed, split, and merged source code files.
In contrast to Beagle, their approach is commit-based and incrementally reconstructs the history based on clone information and file name similarity.
However, the proposed origin analysis is limited to files and thus does not support the tracking of statements and program elements at a more fine-grained level.

Lee et al.~\cite{Tempura:ICSE:2015} implemented a tool named Tempura, enabling code completion and navigation to operate on multiple revisions of code at a time.
To support these features, Tempura pre-processes the commit history of a Git repository, and for each added, modified, renamed, or deleted Java file 
extracts and records its API information (i.e., type, method, field declarations) indexed by the enclosing type's fully qualified name.
Temporal navigation is performed by a simple index lookup to list the revisions in which the selected program element changed.
A major limitation of Tempura is that it requires pre-processing and indexing the repository under analysis, which can take several minutes, especially for large repositories.
Moreover, Tempura is not fully refactoring-aware, as it infers only Class Rename and Move refactorings by leveraging Git's file rename/move detection capability.

Hora et al.~\cite{Hora:ICSE:2018} introduced the concept of \textit{change graph} to model the evolution of classes, methods, and their related changes in the commit history of a project, and study the phenomenon of \textit{untracked} changes.
In this graph, each class or method is represented as a node, while each tracked or untracked change is represented as an edge between two nodes.
However, Hora et al.'s change graph is limited in modeling only the evolution of classes and methods, supports a limited number of refactoring types (5 class-level and 6 method-level refactorings), and uses RefDiff~\cite{Silva:2017} for the detection of refactoring operations, which has inferior precision, recall, and performance than RefactoringMiner~\cite{Tsantalis:ICSE:2018:RefactoringMiner,Tsantalis:TSE:2020:RefactoringMiner2.0}.
Finally, the graph edges model only a small subset of refactoring operations, while other kinds of changes, such as method body and signature changes are omitted.
Thus, Hora et al.'s change graph cannot be used to find all commits where a program element changed, i.e., the graph can provide only the commits in which a program element is involved in refactorings.

Jodavi and Tsantalis~\cite{Jodavi:2022} developed the initial version of CodeTracker, which supports the tracking of method and variable declarations with remarkably high precision and recall (over 99.7\%).
Their work was the first to tackle the program element tracking problem in a fully refactoring-aware fashion
and introduced heuristics for performing \textit{partial} and \textit{incremental} commit analysis to reduce the execution time.
Our current work builds upon the initial version of CodeTracker to solve a way more challenging problem, namely tracking code blocks.
\section{Conclusions and Implications}
In this work, we presented the newer version of our source code tracking tool, CodeTracker 2.0,
which is currently the only tool that can construct the commit change history of code blocks in a fully refactoring-aware fashion.
Moreover, our tool can track code blocks transformed to a different AST type (e.g., \texttt{for} changed to \texttt{while} loop),
and supports forks in the evolution history of a block occurring when two or more different blocks are merged into one.
CodeTracker can extract the complete change history of a code block with a precision and recall of 99.5\% within 3.6 seconds on average and 2 seconds on median.
Finally, a comparison with a baseline based on the GumTree AST diff tool, showed that CodeTracker has +11\% in precision and +21\% in recall over the baseline, and a faster execution time.
\header{Implications for developers}

As illustrated in Section~\ref{sec:chrome-extension-answer-questions}, our tool can potentially help developers understand when, how, and by whom a specific block of code has been changed or introduced.
This is useful for new developers joining a project who want to get familiar with past design choices/changes in the evolution history of an existing software system.
More importantly, our tool supports the migration of code blocks to new language features, such as migrating loops to Java Steam API pipelines, merging \texttt{catch} blocks using the union type for the handled exception types, as well as the conversion of blocks to another block type, such as iterator-based \texttt{while} loops to \texttt{enhanced-for} loops and \texttt{if-else-if} to \texttt{switch} cases.
This feature helps avoid interruptions in the change history due to language migrations and block-type conversions while enabling the tracking of code blocks to their initial introduction, which might have occurred several years ago.
Further user studies must be conducted to evaluate the practical usefulness of the tool in an industrial setting, and assess to what extent the tool supports the actual needs of the developers.


Our tool models change history as a graph allowing forks, and thus it can be used to track the individual change history of merged code blocks.
This feature is particularly useful for clone evolution analysis~\cite{6747168}; when developers want to verify if some duplicated code fragments evolved consistently before they were merged.

\header{Implications for researchers}

Our tool enables several research directions related to Mining Software Repositories (MSR).
For example, researchers can investigate how new language features are getting adopted by developers~\cite{Mazinanian:2017:UUL:3152284.3133909, 10.1145/2568225.2568295, JavaGenericsAdoption},
how and why developers change code blocks to a different block type~\cite{MiningSemanticLoopIdioms},
or extract migration change patterns that could be used to automate similar migrations in other projects.


Finally, our change oracle~\cite{CodeTrackerOracle} can be used to evaluate the accuracy of novel program element tracking tools that will be developed in the future.

\header{Implications for educators}

Our change oracle~\cite{CodeTrackerOracle} includes real-world examples of changes and migrations applied on code blocks that could be used for educating software engineering students and novice developers about the evolution of programming languages and how open-source projects adapt their codebase to new language features.
Each change in our oracle includes the commit, file path, and exact lines in the file where it took place.

\ifCLASSOPTIONcompsoc

\section*{Acknowledgments}
\else
\section*{Acknowledgment}
\fi
This research was partially supported by NSERC grants RGPIN-2018-05095 and RGPIN-2024-04302.

\bibliographystyle{IEEEtran}
\bibliography{IEEEabrv,bibfile}


\end{document}